\documentclass[pdftex,aps,prb,twocolumn,groupedaddress,showpacs,amsmath,amssymb]{revtex4-1}

\usepackage[dvips]{graphicx}            
\usepackage{pslatex}
\usepackage{dsfont}
\usepackage{bm}

\begin{document}


\title{Electrically tunable multiple Dirac cones in thin films of (LaO)$_2$(SbSe$_2$)$_2$ family of materials}


\author{Xiao-Yu Dong$^{1,2\dag}$, Jian-Feng Wang$^{1,2\dag}$, Rui-Xing Zhang$^2$, Wen-Hui Duan$^1$, Bang-Fen Zhu$^1$, Jorge Sofo$^{2}$ and Chao-Xing Liu$^{2\ast}$}
\affiliation{$^{1}$Department of Physics and State Key Laboratory of Low-Dimensional Quantum Physics, Tsinghua University, Beijing 100084, China; }
\affiliation{$^{2}$Department of Physics, The Pennsylvania State University, University Park, Pennsylvania 16802-6300, USA;}
\affiliation{$^\dag$ These authors contributed equally to this work;}
\affiliation{$^\ast$ e-mail:  cxl56@psu.edu}

\date{\today}

\begin{abstract}
Two-dimensional Dirac physics has aroused great interests in condensed matter physics ever since the discovery of graphene and topological insulators. The ability to control the properties of Dirac cones, such as bandgap and Fermi velocity, is essential for various new phenomena and the next-generation electronic devices. Based on first-principles calculations and an analytical effective model, we propose a new Dirac system with eight Dirac cones in thin films of the (LaO)$_2$(SbSe$_2$)$_2$ family of materials, which has the advantage in its tunability: the existence of gapless Dirac cones, their positions, Fermi velocities and anisotropy all can be controlled by an experimentally feasible electric field. We identify layer dependent spin texture induced by spin-orbit coupling as the underlying physical reason for electrical tunability of this system. Furthermore, the electrically tunable quantum anomalous Hall effect with a high Chern number can be realized by introducing magnetization into this system.
\end{abstract}


\maketitle

\section{Introduction}\label{sec:intro}
Unlike the usual materials with parabolic energy dispersion described by the non-relativistic Schr$\ddot{o}$dinger equation, graphene\cite{neto2009} provides the first example in condensed matter physics with low energy effective physics described by the relativistic Dirac equation with linear energy dispersion. Later, it was realized that two dimensional Dirac type of dispersion also appears for the surface states of three dimensional topological insulators (TIs)\cite{hasan2010,qi2011}, of which spin is resolved and locked to the momentum, forming a spin texture in the surface Brillouin zone (BZ). Dirac Hamiltonians with or without mass also exist in the low energy physics of several other families of materials, including topological crystalline insulators (such as SnTe\cite{hsieh2012,tanaka2012,dziawa2012}), group-VI dichalcogenides (such as MoS$_2$\cite{splendiani2010,mak2010,xiao2012,cao2012,zeng2012}), SrMnBi$_2$\cite{park2011}, $d$-wave cuprate superconductors\cite{tsuei2000}, and three-dimensional Dirac semimetals (such as Na$_3$Bi\cite{liu2014}, Cd$_3$As$_2$\cite{borisenko2013,neupane2014}). Due to the similar energy dispersion, all these materials share some common and uniqpue physical properties, thus dubbed Dirac materials\cite{wehling2014}, which are believed to have potential applications in high performance nanoelectronics\cite{neto2009}, spintronics, and quantum computation\cite{hasan2010,qi2011}. In these existing Dirac materials, the properties of Dirac cones, such as the position and Fermi velocity, are usually determined by intrinsic properties of material, such as crystal structures and spin-orbit coupling (SOC), and thus difficult to be controlled experimentally. For example, pristine graphene\cite{neto2009} does not have a bandgap, so for the potential application in transistors, one needs bilayer graphene\cite{zhang2009a,xia2010}, of which the energy dispersion becomes parabolic. Therefore, it is desirable to have Dirac materials with the properties of Dirac cones tunable by external experimental conditions.

In this paper, we propose a new family of Dirac materials, including ($R$O)$_2$($MX_2$)$_2$\cite{guittard1984,yazici2013,mizuguchi2012,demura2013} and ($Ae$F)$_2$($MX_2$)$_2$\cite{kabbour2006,lei2013,lin2013} ($R$ = rare earth, $Ae$ = Sr or Ba, $M$ = Sb or Bi, $X$ = S, Se or Te, O = oxygen and F = fluorine). The multiple Dirac cones and the electrical tunability of Dirac physics in this system allow us to obtain the quantum anomalous Hall effect with higher Chern number, which can be controlled by a gate voltage.
\section{Electrically tunable Dirac cones}
To illustrate crystal structures of this family of materials, we may take (LaO)$_2$(SbSe$_2$)$_2$ as an example, which possesses the tetragonal ZrCuSiAs-type structure\cite{johnson1974} with space group $P4/nmm$. As shown in Fig. 1a, (LaO)$_2$(SbSe$_2$)$_2$ has a triple-layer (TL) structure with one (LaO)$_2$ layer sandwiched by two SbSe$_2$ layer in one unit cell. The (LaO)$_2$ layer is similar to that in LaOFeAs\cite{stewart2011}, formed by a square lattice of O atoms, coordinated tetrahedrally with four neighboring La atoms. However, the SbSe$_2$ layer has a different structure from the FeAs layer. For one SbSe$_2$ layer, the Sb$_1$ (Sb$^{\prime}_1$) atom and Se$_2$ (Se$^{\prime}_2$) atom form a distorted checkerboard lattice while the Se$_1$ (Se$^{\prime}_1$) atom lies between (LaO)$_2$ layer and the checkerboard SbSe layer. Each Sb atom can be viewed to have a distorted octahedral coordination with six Se atoms. Strong ionic bonding is formed between the central (LaO)$_2$ layer and two adjacent SbSe$_2$ layers within one unit cell while the chemical bonding between two adjacent SbSe$_2$ layers in neighboring unit cells is much weaker, dominated by van der Waals interaction\cite{liu2013}. The primitive lattice vectors are shown in Fig. 1a, denoted as $\mathbf{x}$, $\mathbf{y}$, $\mathbf{z}$. The lattice constant is denoted as $a$ along the $x$ and $y$ directions and $c$ along the $z$ direction. The space group $P4/nmm$ is non-symmorphic and possess a glide operation $\hat{g}_z=\{\hat{m}_z|\bm{\tau}\}$ which consists of a reflection in the $xy$-plane, $\hat{m}_z:(x,y,z)\rightarrow(x,y,-z)$, followed by a non-primitive translation $\bm{\tau}=(\frac{a}{2},\frac{a}{2},0)$. The system has inversion symmetry $\hat{I}$ with the  inversion center located at the center of two nonequivalent O atoms. In addition, there is a mirror symmetry along both the $x$ and $y$ directions, denoted as $\hat{m}_x$ and $\hat{m}_y$.
\begin{figure} 
\includegraphics[width=8.6cm]{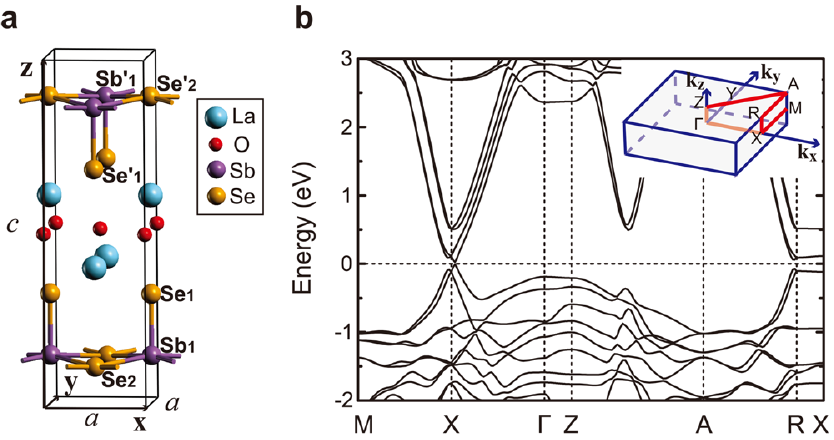}
\caption{{\bf Crystal structure and bulk band dispersion.} {\bf a}, Primitive cell of (LaO)$_2$(SbSe$_2$)$_2$ crystal structure. {\bf b}, Band structure of (LaO)$_2$(SbSe$_2$)$_2$ along the high symmetric lines which are depicted by red lines in bulk BZ (the inset).} \label{Fig1}
\end{figure}
Based on the density-functional theory, we calculate the electronic band structure of (LaO)$_2$(SbSe$_2$)$_2$, which is found to be a narrow gap semiconductor with the bandgap near the X and Y points in the BZ, as shown in Fig. 1b. Both conduction and valence bands are four-fold degenerate at X
and split into two doubly degenerate states along the $\Gamma$--X line. An anti-crossing occurs between the conduction and valence bands along the $\Gamma$--X line, resulting in a bandgap around 20 meV. In addition, the band structure is found to be non-dispersive along the line $\Gamma$--Z and R--X, indicating that adjacent TLs are weakly coupled and the system is essentially two dimensional. Therefore, we will study a film configuration with one TL of (LaO)$_2$(SbSe$_2$)$_2$. It has recently been shown that a TL film is stable for one material (LaO)$_2$(BiS$_2$)$_2$ in this family\cite{liu2013}. The configuration of a TL film is shown schematically in Fig. 2a, with front and back gate voltages $V_{\text{g}}$, which provide an electric field $E_{\mathrm{ex}}$ or an effective asymmetric potential e$U$ on the thin film. Fig. 2b--e show band dispersion under the electric field $E_{\mathrm{ex}}=0,9.2,18.4,27.6$ mV {\AA}$^{-1}$, respectively. With increasing electric fields, the double degeneracy along the $\Gamma$--X line is split, and the bandgap is reduced and closed around $E_{\mathrm{ex}}=9.2$ mV {\AA}$^{-1}$. From Fig. 2c--e, one can clearly see that for the electric field $E_{\mathrm{ex}}\geq9.2$ mV {\AA}$^{-1}$, the bandgap will re-open along the X--M line while it remains gapless along the $\Gamma$--X line. There are totally eight gapless points in the whole BZ, four around X and four around Y, and their positions in the BZ are indicated by red dots in the inset of Fig. 2c. Each gapless point forms a Dirac cone with linear dispersion, as shown in the insets of Fig. 2d and e. These eight Dirac cones can be classified into two sets: four equivalent Dirac cones (A) are close to X or Y while the other four (B) are away from these two momenta. We emphasize that all the Dirac cones here are spin-resolved, different from that in graphene but similar to the case of TIs. Strikingly, the properties of Dirac cones, such as their positions, velocities and anisotropy, are tunable with electric fields. For example, the velocity of Dirac fermions around X is shown as a function of the electric field $E_{\mathrm{ex}}$ in Fig. 2f, which can be tuned in the range $(4.6\sim6.7)\times 10^{5}$ m s$^{-1}$ along the $k_x$ direction and $(0.2\sim4.0)\times 10^{5}$ m s$^{-1}$ perpendicular to $k_x$ with the experimentally feasible electric field\cite{novoselov2004,castro2007} [the corresponding voltage drops between the two checkerboard SbSe layers are in the range $0\sim400$ mV (see Supplementary Figure 1 and Supplementary Note 1 for the relationship between the electric field and the atomic layer energy splitting)]. We notice that the velocity of linear dispersion has a small anisotropy for $E_{\mathrm{ex}}<18.4$ mV {\AA}$^{-1}$, while the velocity anisotropy increases rapidly when $E_{\mathrm{ex}}\geq18.4$ mV {\AA}$^{-1}$, as clearly shown in the insets of Fig. 2d and e. At the electric field $E_{\mathrm{ex}}=36.8$ mV {\AA}$^{-1}$, the velocity along the $\Gamma$--X direction is around $7\times 10^5$ m s$^{-1}$, about 4 times smaller than that in graphene, while that perpendicular to the $\Gamma$--X direction is one order smaller ($2\times 10^4$ m s$^{-1}$). To understand the existence and tunability of Dirac cones, we will next develop a low energy effective model for this system.

\begin{figure} 
\includegraphics[width=8.6cm]{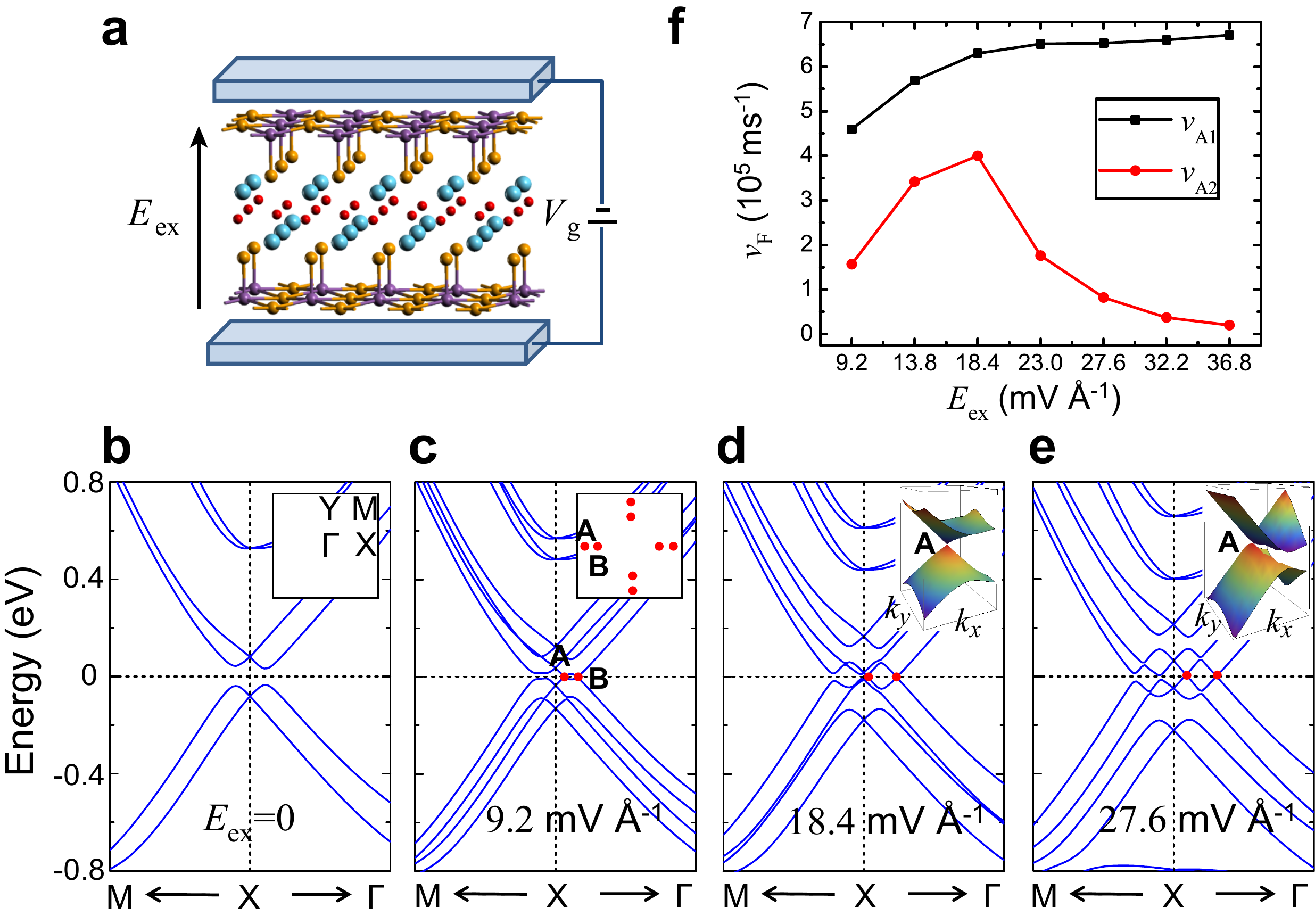}
\caption{{\bf Electric tunability of Dirac cones.} {\bf a}, Schematic plot of a TL film of (LaO)$_2$(SbSe$_2$)$_2$ with a gate voltage $V_\text{g}$ or an electric field $E_{\mathrm{ex}}$. {\bf b}--{\bf e}, Band dispersions of a TL film near X under the electric fields $E_{\mathrm{ex}}=0,9.2,18.4,27.6$ mV {\AA}$^{-1}$, respectively. The inset of {\bf b} shows the BZ and inset of {\bf c} shows the positions of the gapless points in the BZ. The red dots depict gapless points. The insets of {\bf d} and {\bf e} show Dirac cones at the red dots close to the X point from the first-principles calculations. The plot range is $0.4957\frac{2\pi}{a}\leq k_x \leq 0.4997\frac{2\pi}{a}$, $-0.002\frac{2\pi}{a}\leq k_y \leq 0.002\frac{2\pi}{a}$, $-0.02$ eV$\leq E \leq 0.02$ eV for {\bf d}, and $0.4831\frac{2\pi}{a}\leq k_x \leq 0.4991\frac{2\pi}{a}$, $-0.008\frac{2\pi}{a}\leq k_y \leq 0.008\frac{2\pi}{a}$, $-0.06$ eV$\leq E \leq 0.06$ eV for {\bf e}. {\bf f}, The Fermi velocity is shown as a function of the electric field $E_{\mathrm{ex}}$. $v_{\text{A1}}$ ($v_{\text{A2}}$) indicates the velocity of Dirac cone A along the direction of $k_x$ ($k_y$).} \label{Fig2}
\end{figure}

\section{Low-energy effective model}
To develop a theoretical model, we first check the orbital natures of conduction and valence bands. The band structure with different atomic projections is shown in Fig. 3a and b. The orbitals of La and O atoms lie far away from the Fermi energy due to the strong electron negativity and affinity. The outmost shells for both Sb ($5s^25p^3$) and Se ($4s^24p^4$) are $p$ orbitals. The Se$_1$ and Se$^{\prime}_1$ atoms are close to the (LaO)$_2$ layer and form strong bonds, which push their energy levels away from the Fermi energy. Thus, the bands near the Fermi energy are dominated by the $p$ orbitals of Sb$_1$ (Sb$^{\prime}_1$) and Se$_2$ (Se$^{\prime}_2$) atoms in the SbSe checkerboard layers. Therefore, we focus on the bilayer SbSe checkerboard lattice and develop an atomic tight-binding (TB) model on this lattice with three $p$ orbitals on each site. The details of lattice structures (Supplementary Figure 2) and the form of TB model are described in Supplementary Note 2. By carefully choosing parameters, we can qualitatively reproduce energy dispersions near X (or Y) and four Dirac cones emerge along the $\Gamma$--X line after turning on an asymmetric potential between two layers, as shown in Fig. 3c and d, Supplementary Figure 3 and Supplementary Note 3.

\begin{figure} 
\includegraphics[width=8.6cm]{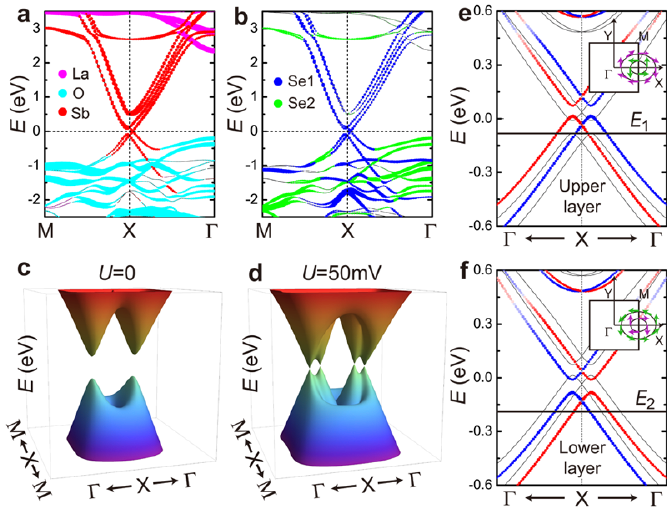}
\caption{{\bf Atomic orbital and spin nature of band structure.} {\bf a}--{\bf b}, Band structures with different atomic projections. The atomic characters La, O, Sb, Se$_1$, and Se$_2$ are indicated by magenta, cyan, red, green, and blue, respectively. {\bf c}--{\bf d}, Band structure around X from TB model, with $U=0$ in {\bf c} and $U=50$ mV in {\bf d}. {\bf e}--{\bf f}, The {\it ab initio} calculations of spin-resolved bands with the upper (in {\bf e}) and lower (in {\bf f}) SbSe checkerboard layer projection under the electric field $E_{\mathrm{ex}}=9.2$ mV {\AA}$^{-1}$. The red and blue denote the states of spin up and spin down which are polarized along the $y$ direction. The two insets schematically reveal the spin textures under the iso-energetic contours with the energy $E_1$ and $E_2$, respectively.} \label{Fig3}
\end{figure}

Below, we will focus on the effective theory around X (or Y). We find that the conduction band minimum and valence band maximum are dominated by $p_x$ ($p_y$) orbitals of Se and Sb atoms around the X (Y) point. This orbital nature is confirmed by the {\it ab initio} calculation, as shown in Supplementary Figure 4 of Supplementary Note 4. As described above, both conduction and valence bands have four-fold degeneracy at X, which originates from two spin states and two checkerboard SbSe layers. Therefore, we can construct a four band effective model around X on the basis $|p_x,\sigma,\xi_\mu\rangle$ for the conduction and valence bands respectively, where $\sigma=\uparrow_z, \downarrow_z$ denotes spin, $\xi=$c,v denotes conduction band and valence band, and $\mu=\pm$ denotes the upper and lower layer. We choose the spin axis along the $z$ direction. It should be emphasized that there is a strong hybridization between Se and Sb atoms for both the conduction and valence bands near the bandgap. Thus, we use $\xi=$c,v instead of Sb,Se to denote the basis. Using the L$\ddot{o}$wdin perturbation method\cite{winkler2003}, we find the effective Hamiltonian takes the form
\begin{eqnarray}
	 H_{\xi=\text{c},\text{v}}(\mathbf{k})&=&\epsilon_{\xi}(k_x,k_y)+f_{1\xi}k_x\hat{\tau}_x+ f_{2\xi}k_x\hat{\sigma}_y\hat{\tau}_z\nonumber\\&&-f_{3\xi}k_y\hat{\sigma}_x \hat{\tau}_z+eU\hat{\tau}_z
	\label{eq:Heff}
\end{eqnarray}
around X up to the second order in $\mathbf{k}$ for the conduction or valence band, where $\epsilon_{\xi}(k_x,k_y)=d_{0\xi}+d_{1\xi}k_x^2+d_{2\xi}k_y^2$, the Pauli matrices $\hat{\tau}$ denote layer index and $\hat{\sigma}$ denote spin index. $d_{i\xi}$ and $f_{i\xi}$ are material dependent parameters, and can be extracted from perturbation procedure. In the Hamiltonian (\ref{eq:Heff}), the first term is not important. The second term is a spin-independent term, describing the hybridization between two layers. The third and fourth terms depend on spin and originate from the third order perturbation combining the interlayer hopping and SOC (see Supplementary Note 5). We notice that these two terms take the familiar form of the Rashba type of Hamiltonian while the additional $\hat{\tau}_z$ dependence indicates that the spin splitting is opposite for two layers. This layer dependent Rashba term is the origin of the layer dependent spin texture described below. The last term describes the asymmetric potential e$U$ induced by electric fields.

We next look at the symmetry properties of Hamiltonian (\ref{eq:Heff}). The present system has time reversal (TR) symmetry $\hat{T}$ and the space group symmetry $P4/nmm$. The wavevector group of $P4/nmm$ at X contains glide symmetry $\hat{g}_z$, mirror symmetry $\hat{m}_y$ and inversion symmetry $\hat{I}$. The effective Hamiltonian (\ref{eq:Heff}) can be derived from symmetry principles based on the above symmetries, as shown in Supplementary Note 6 and Supplementary Table 1. We find that the four-fold degeneracy at X can be determined from three symmetry operations $\hat{T}$, $\hat{g}_z$ and $\hat{m}_y$. On the basis $|\sigma,\mu\rangle$ of the effective Hamiltonian (we neglect $p_x$ and $\xi$ for short), the representation of symmetry operators is given by $g_z=\hat{\sigma}_z\hat{\tau}_x$, $m_y=i\hat{\sigma}_y$ and $T=+(-)i\hat{\sigma}_y\hat{\tau}_zK$ for conduction (valence) bands, where $K$ is complex conjugate. Since $[m_y,H_{\xi}]=0$ at X, we may choose eigen states with definite mirror parity of $\hat{m}_y$. Four eigen states can be written as $|\psi_{\mu,\uparrow_y(\downarrow_y)}\rangle=\frac{1}{\sqrt{2}} (|\uparrow_z,\mu\rangle+(-)i|\downarrow_z,\mu\rangle)$, which satisfy $\hat{m}_y|\psi_{\mu,\uparrow_y(\downarrow_y)}\rangle=+i(-i) |\psi_{\mu,\uparrow_y(\downarrow_y)}\rangle$. Here $\uparrow_y$ and $\downarrow_y$ correspond to up and down spin along the $y$ direction. Now let us consider how $\hat{T}$ and $\hat{g}_z$ act on these four states $|\psi_{\mu,\sigma_y}\rangle$. Both $\hat{T}$ and $\hat{g}_z$ change spin up $\uparrow_y$ to spin down $\downarrow_y$. However, the obtained states operated by $\hat{T}$ or $\hat{g}_z$ are different since the layer index is preserved by $\hat{T}$ but changed by $\hat{g}_z$. In addition, the combination of $\hat{T}$ and $\hat{g}_z$ leads to the state with the same spin but different layer indices. Thus, these four eigen states can be related to each other by $\hat{T}$, $\hat{g}_z$ and $\hat{T}\hat{g}_z$, so they must be degenerate at X. In general, the four-fold degeneracy at the X (Y) point is a direct consequence of the combination of TR symmetry $\hat{T}$ and the anti-commutation relation between $\hat{g}_z$ and $\hat{m}_y$ ($\hat{m}_x$).

After understanding the degeneracy at X, we next consider the states away from X along the $\Gamma$--X line ($k_x\neq 0, k_y=0$ in Hamiltonian (\ref{eq:Heff})). In Fig. 2b, one can see that four-fold degenerate states are split into two doubly degenerate states. According to Hamiltonian (\ref{eq:Heff}), both the hybridization term ($f_{1\xi}$ term) and the SOC term ($f_{2\xi}$ and $f_{3\xi}$ terms) can contribute to this splitting. The remaining double degeneracy at a finite $k_x$ comes from the combined symmetry $\hat{T}\hat{I}$ (Kramers' doublet due to both the TR symmetry and inversion symmetry). Alternatively, we can also understand it from the combination of $\hat{g}_z$ and $\hat{m}_y$.
We can still choose two degenerate eigen-states to have definite mirror parity, denoted as $m_y|\psi_{\uparrow_y(\downarrow_y)}(k_x)\rangle=+i(-i)|\psi_{\uparrow_y(\downarrow_y)}(k_x)\rangle$. It should be emphasized that the states $|\psi_{\uparrow_y(\downarrow_y)}(k_x)\rangle$ should be a linear combination of the basis in different layers. $|\psi_{\uparrow_y}(k_x)\rangle$ and $|\psi_{\downarrow_y}(k_x)\rangle$ are still related to each other by $\hat{g}_z$ which will also change the layer index. This means that if $|\psi_{\uparrow_y}(k_x)\rangle$ mainly stays at the upper layer, $|\psi_{\downarrow_y}(k_x)\rangle$ must be at the lower layer. Thus, spin and layer indices are related to each other, leading to the layer dependence of spin textures in this system. To show it more explicitly, we calculate spin polarization at a given momentum $\mathbf{k}$ for different layers based on the effective model (\ref{eq:Heff}). As an example, we may consider spin texture of two degenerate eigen-states $|\psi^{\xi}_{\alpha,+}\rangle$ ($\alpha=1,2$) with eigen-energy $E_+=\epsilon_\xi+\sqrt{(f_{1\xi}^2+f_{2\xi}^2)k_x^2 +f_{3\xi}^2k_y^2}$, which is given by
\begin{eqnarray}
	&&\langle {\bf S}_\mu\rangle_{\xi}=\sum_{\alpha} \langle \psi^{\xi}_{\alpha,+}|{\bf S}_\mu|\psi^{\xi}_{\alpha,+}\rangle\nonumber\\&&=\frac{\mu\hbar} {2\sqrt{(f_{1\xi}^2+f_{2\xi}^2)k_x^2+f_{3\xi}^2k_y^2}}(- f_{3\xi}k_y\hat{\bf x}+f_{2\xi}k_x\hat{\bf y}),
	\label{eq:Spintexture}
\end{eqnarray}
where the layer dependent spin operator is defined as $S_{i,\mu}=\frac{\hbar}{2}\hat{\sigma}_i\frac{\hat{\tau}_z+ \mu}{2}$ with $i=x,y,z$. 
From Eq. (\ref{eq:Spintexture}), we indeed find opposite spin textures for different layers once the parameter $f_{2\xi}$ or $f_{3\xi}$, originating from SOC, is non-zero.

Now let us turn on the asymmetric potential e$U$ between two layers.  Since spin states are locked to the layer indices, two spin states with opposite layer indices will be split accordingly. This picture has been utilized to explain the giant Rashba spin splitting in (LaO)$_2$(BiS$_2$)$_2$ in the early study\cite{liu2013,zhang2014a}. In our case, we need to consider spin splitting for both conduction and valence bands. It is found that the parameters $f_{2\text{c}}$ and $f_{2\text{v}}$ have opposite signs ($f_{2\text{c}}>0$ and $f_{2\text{v}}<0$) while $f_{3\text{c}}$ and $f_{3\text{v}}$ have the same sign ($f_{3\text{c}},f_{3\text{v}}<0$), leading to different spin textures for the conduction and valence bands. From the Supplementary Note 7 and Supplementary Figure 5, for the case with $U>0$, $k_x>0$ and $k_y=0$, the lowest conduction band carries spin up at the lower layer while the highest valence band is dominated by the states with spin down at the upper layer. Since these two states have opposite spin, as well as opposite mirror parities of $\hat{m}_y$, there is no coupling between them. Thus, the Dirac cones due to the crossing points between the conduction and valence bands along the $\Gamma$-X line are protected by mirror symmetry. From the above analysis, we can see that the layer dependent spin texture is the underlying physical reason for the existence and tunability of Dirac cones. To further confirm our physical picture, we perform the first-principles calculations of layer dependent spin texture in this system, as shown in Fig. 3e and f. Fig. 3e (f) is the spin-resolved band projecting on the upper (lower) SbSe checkerboard layer at the electric field $E_{\mathrm{ex}}=9.2$ mV {\AA}$^{-1}$. The opposite spin textures for the upper and lower SbSe layer are shown in the insets of Fig. 3e and f, respectively.

\section{Electrically tunable quantum anomalous Hall effect}
When gapless Dirac cones are gapped by magnetization, quantized Hall conductance is possible to arise even at a zero magnetic field and without Landau levels. This phenomenon is known as the quantum anomalous Hall effect\cite{haldane1988,onoda2003,qi2006,liu2008,yu2010}, which was recently observed in Cr or V doped (Bi,Sb)$_2$Te$_3$ films experimentally with the quantized Hall conductance $\frac{e^2}{h}$ (Chern number 1)\cite{chang2013,chang2015}. Multiple-Dirac-cones systems allow for the realization of the quantum anomalous Hall effect with a larger Chern number, which was first proposed in graphene systems\cite{qiao2010}, as well as other systems, including magnetically doped (Bi,Sb)$_2$Te$_3$ films\cite{jiang2012} and SnTe systems\cite{fang2014}. Here we will consider the exchange coupling of magnetic moments and predict the quantum anomalous Hall (QAH) effect with a high Chern number in magnetically doped (LaO)$_2$(SbSe$_2$)$_2$, which can be controlled by an external electric field. To theoretically study how exchange coupling affects our system, we construct a more realistic TB model using the maximum localized Wannier function method\cite{marzari1997,souza2001}, and then introduce exchange coupling phenomenologically. This method has been widely adopted to study the QAH effect in magnetically doped TIs\cite{yu2010,wang2013}.
\begin{figure} 
\includegraphics[width=8.6cm]{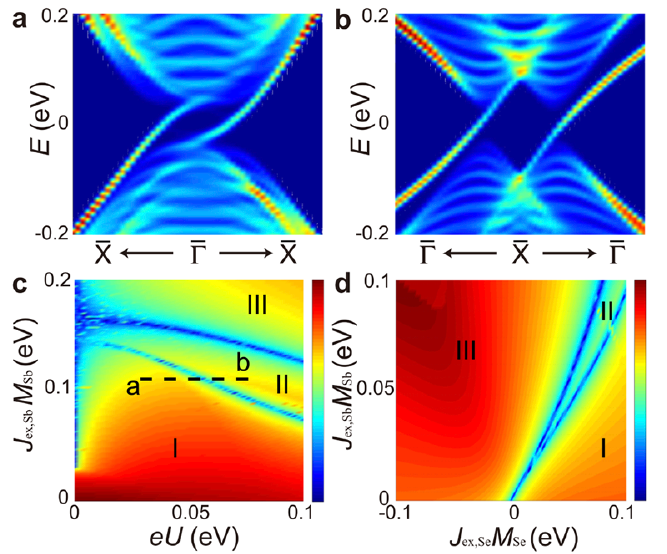}
\caption{{\bf Electrically tunable quantum anomalous Hall effect.} The local density of states at one edge of a ribbon of (SrF)$_2$(SbSe$_2$)$_2$ film near the $\bar{\Gamma}$ point in {\bf a} and near the $\rm{\bar{X}}$ point in {\bf b} for $eU=0.1$ eV and $J_{\text{ex},\text{Se}}M_{\text{Se}}=-J_{\text{ex}, \text{Sb}}M_{\text{Sb}}=0.05$ eV. Two chiral edges are found near $\bar{\Gamma}$ while the other two near $\rm{\bar{X}}$. The bulk bandgap is plotted as a function of $J_{\text{ex},\text{Sb}}M_{\text{Sb}}$ and $eU$ in {\bf c} for $J_{\text{ex},\text{Se}}M_{\text{Se}}=0.1$ eV, while it is shown as a function of $J_{\text{ex},\text{Sb}}M_{\text{Sb}}$ and $J_{\text{ex},\text{Se}}M_{\text{Se}}$ in {\bf d} for $eU=0.05$ eV. Here the gaps are plot in logarithmic scale and blue color is for gap closing. We find two gapless lines (blue lines) divide the phase diagram into three regimes I, II and III with Hall conductance $\frac{4e^2}{h}$, $0$ and $-\frac{4e^2}{h}$, respectively. By tuning the electric field the states can move along the dashed line from a to b.}   \label{Fig4}
\end{figure}

We start from the case with an asymmetric potential e$U$$=100$ meV and no magnetization, in which eight Dirac cones exist in the whole BZ. The results obtained from the Wannier function method is consistent with the direct calculation from the first-principles methods. The exchange coupling is introduced into the calculation by the phenomenological Kondo-type of Hamiltonian $H_{\text{ex}}=J_{\text{ex}}\mathbf{s}\cdot\mathbf{M}$, where $J_{\text{ex}}$ is the exchange coupling constant, $\mathbf{s}$ denotes electron spin and $\mathbf{M}$ is the average magnetization of the system.
The Dirac cones are gapped by turning on magnetization (see Supplementary Figure 6 in Supplementary Note 8).
It is well-known that each gapped 2D Dirac cone (massive Dirac Hamiltonian) contributes half quantized Hall conductance ($\frac{e^2}{2h}$ or $-\frac{e^2}{2h}$). Thus, the total Hall conductance is determined by the sign of the Hall conductance contribution from different Dirac cones. It turns out that all massive Dirac Hamiltonians take the same sign, leading to the total Hall conductance $\frac{4e^2}{h}$ or $-\frac{4e^2}{h}$ for the whole system. To see this more explicitly, we directly calculate edge states of the whole system in a ribbon configuration and plot the local density of states at one edge along the ${\bf x}$ direction. We take (SrF)$_2$(SbSe$_2$)$_2$ as an example due to its large electric tunability. Similar results can also be obtained for other systems in this class of materials, as shown in the Supplementary Note 8 and Supplementary Figure 7. As shown in Fig. 4a and b, there are in total four chiral edge states propagating along the same direction, with two appearing near the $\rm{\bar{X}}$ point and the other two near the $\bar{\Gamma}$ point, where $\bar{\Gamma}$ and $\rm{\bar{X}}$ are the projection of $\Gamma$ and X into the 1D edge. Therefore, the edge state picture is consistent with the analysis of bulk Dirac cones, revealing that the QAH state with the Hall conductance $\pm\frac{4e^2}{h}$ can be realized in this system.

Fig. 4c and d show bulk bandgap as a function of magnetization ($J_{\text{ex},\text{Sb}}M_{z,\text{Sb}}$ and $J_{\text{ex},\text{Se}}M_{z,\text{Se}}$) and asymmetric potential e$U$. Since a topological phase cannot be changed when the bulk bandgap remains open, the phase diagram can be determined by tracking the gap closing lines\cite{liu2008}. Two gapless lines are found in Fig. 4c, indicating two topological phase transitions. For each metallic line, four equivalent Dirac cones (Dirac cones A or B in the inset of Fig. 2c) reverse their bandgap, leading to the change of Hall conductance by $\frac{4e^2}{h}$. Therefore, we can determine the Hall conductance in each regime, as dictated by the regimes I, II and III in Fig. 4c with Hall conductance $\frac{4e^2}{h}$, $0$ and $-\frac{4e^2}{h}$, respectively. We emphasize that along the dashed line $a-b$ in Fig. 4c, a topological phase transition from the Hall conductance $\frac{4e^2}{h}$ to $0$ can occur by tuning only electric fields and fixing magnetization. This again reflects the electrical tunability of Dirac physics in this system, as discussed in the above section.

\section{Discussion and conclusion}
The physics discussed above for (LaO)$_2$(SbSe$_2$)$_2$ can also be applied to other materials in this family. Since all low energy physics occurs in the SbSe$_2$ layer, the (LaO)$_2$ layer can also be replaced by other ($R$O)$_2$ layer where $R$ is a rare earth atom. As shown in Supplementary Note 4 and Supplementary Figure 8, the SbSe$_2$ layer can also be replaced by other Sb$X_2$ layers ($X$ = Te, S). For Bi$X_2$ ($X$ = S,Te,Se)\cite{usui2012,shein2013}, an indirect bandgap occurs, ranging from 70 to 300 meV, and is not suitable for electric control. In addition, the (LaO)$_2$ layer can also be replaced by ($Ae$F)$_2$ where $Ae$ = Sr, Ba \cite{kabbour2006,lei2013,lin2013}. Experimentally, bulk (LaO)$_2$(SbSe$_2$)$_2$ and ($Ae$F)$_2$(SbSe$_2$)$_2$ have been fabricated by the high-temperature ceramic method\cite{guittard1984,kabbour2006}. As shown in Supplementary Note 9 and Supplementary Figure 9, the binding energy between the TLs for (LaO)$_2$(SbSe$_2$)$_2$ is about 7.3 meV {\AA}$^{-2}$ without van der Waals (vdW) correction and 30.7 eV {\AA}$^{-2}$ with vdW correction, comparable to that of MoS$_2$ ($\sim20$ meV {\AA}$^{-2}$ )\cite{bjorkman2012}. The exfoliation of TL (LaO)$_2$(SbSe$_2$)$_2$ is expected to be feasible in experiments. For the case of multiple TLs, Dirac cones also exist but may be buried in other bulk bands, as shown in the Supplementary Note 10 and Supplementary Figure 10. The effect of strain (Supplementary Figure 11 in Supplementary Note 11) and lattice distortion (Supplementary Figure 12 and 13 in Supplementary Note 9) and the electrical tunability by gate voltage (Supplementary Figure 1 in Supplementary Note 1) are also carefully examined, from which we find Dirac physics is quite robust for a single TL film. Magnetic moments can be introduced into this system by magnetic doping, which have been successfully used for TI materials, such as Sb$_2$Te$_3$ family of materials, to realize the QAH effect\cite{chang2013}. Alternatively, one can also exfoliate the TL of (LaO)$_2$(SbSe$_2$)$_2$ and transfer it to ferromagnetic substrate to induce exchange coupling by ferromagnetic proximity\cite{wei2013}. A similar technique has been applied to the graphene systems to observe the anomalous Hall effect\cite{wang2015,qiao2010}.

The electrical tunability of Dirac physics in this system indicates its potential application in various fields. For example, we have shown that a topological phase transition between the QAH states with different quantized Hall conductance can be achieved by controlling electric fields, which will be useful for the experimental study of critical phenomena of topological phase transitions\cite{chang2013,checkelsky2014}. Moreover, superconductivity has been realized in (LaO)$_2$(BiS$_2$)$_2$ \cite{mizuguchi2012,yazici2013,demura2013,lin2013} or in LaOFeAs, a well-studied unconventional superconductor with a similar crystal structure\cite{kamihara2008}. Therefore, it is possible to fabricate heterostructures combining (LaO)$_2$(SbSe$_2$)$_2$ and these superconducting materials, which provide a new flatform to study the coexistence of Dirac physics and superconductivity.

\section{Computational methods}
All the first-principles calculations are based on the density-functional theory as implemented in the Vienna \textit{ab initio} simulation package (VASP)\cite{kresse1996A, kresse1996B}. The projector augmented wave method\cite{blochl1994} is used, with a kinetic energy cutoff of 400 eV for the plane wave basis set. The generalized gradient approximation of the Perdew-Burke-Ernzerhof (PBE) type functional\cite{perdew1996} is adopted to describe the exchange-correlation interaction. In the crystal structure, we take the experimental lattice constants: $a$ = 4.13 {\AA}, and $c$ = 14.17 {\AA} \cite{guittard1984}. For the TL thin film, we build a slab model with a vacuum region of 14 {\AA} to decouple the consecutive slabs in the supercell approach. 13$\times$13$\times$5 and 13$\times$13$\times$1 $\Gamma$-centered $k$-point meshes are used in the bulk and slab calculations, respectively. The SOC is employed in all electronic structure calculations. We also perform electronic band calculations using the WIEN2K package\cite{blaha2001} and reproduce energy dispersion obtained by VASP. The $18\times 18$ tight binding Hamiltonian, which nicely captures the low energy physics around the Fermi level, is constructed using the maximal localized Wannier function method\cite{marzari1997,souza2001}. We choose $p_x$, $p_y$, $p_z$ states of Sb and Se atoms as the projection centers.

\begin{acknowledgments}
We would like to thank Binghai Yan for useful discussions and Gang Yang for the help of the WIEN2K package. CXL acknowledge the support from Office of Naval Research (Grant No. N00014-15-1-2675). XYD and BFZ acknowledge the support from National Natural Science Foundation of China (Grant No. 11374173). XYD, JFW, WHD and BFZ acknowledge the Program of Basic Research Development of China (Grant No. 2011CB921901). JFW and WHD acknowledge the support from National Natural Science Foundation of China (Grant No. 11334006).
\end{acknowledgments}

\newpage

\renewcommand\thesection{{\bf Supplementary Note \arabic{section}}}
\renewcommand{\figurename}{{\bf Supplementary Figure}}
\renewcommand{\tablename}{{\bf Supplementary Table}}
\setcounter{section}{0}
\setcounter{table}{0}
\setcounter{figure}{0}
\begin{widetext}
\section{Relationship between the applied electric field and the atomic layer energy splitting}
In the article, we applied the external electric field $E_{\mathrm{ex}}$ to the TL slab in the DFT calculations, and used the atomic layer energy splitting e$U$ in the tight-binding model and low-energy effective model analysis. To build the link between them and confirm the predicted phases are really available for experimentally feasible gate voltages, here we give the relationship between the applied electric field $E_{\mathrm{ex}}$ (as well as the corresponding voltage drop between the two SbSe layers) and the atomic layers energy splitting e$U$ based on the DFT calculations. From Supplementary Figure~\ref{FigS12}, we can see that energy splitting e$U$ between the two SbSe atomic layers is slightly different from the voltage drop for the valence band maximum (VBM). The difference of the conduction band minimum (CBM) is even larger as the applied electric field increases. Since we add an external electric field in our calculation, the screening effect between two SbSe layers has automatically been taken into account. We expect the screening is quite weak since we only consider a single TL (LaO)$_2$(SbSe$_2$)$_2$.

\begin{figure} [h]
\includegraphics[width=0.6\textwidth]{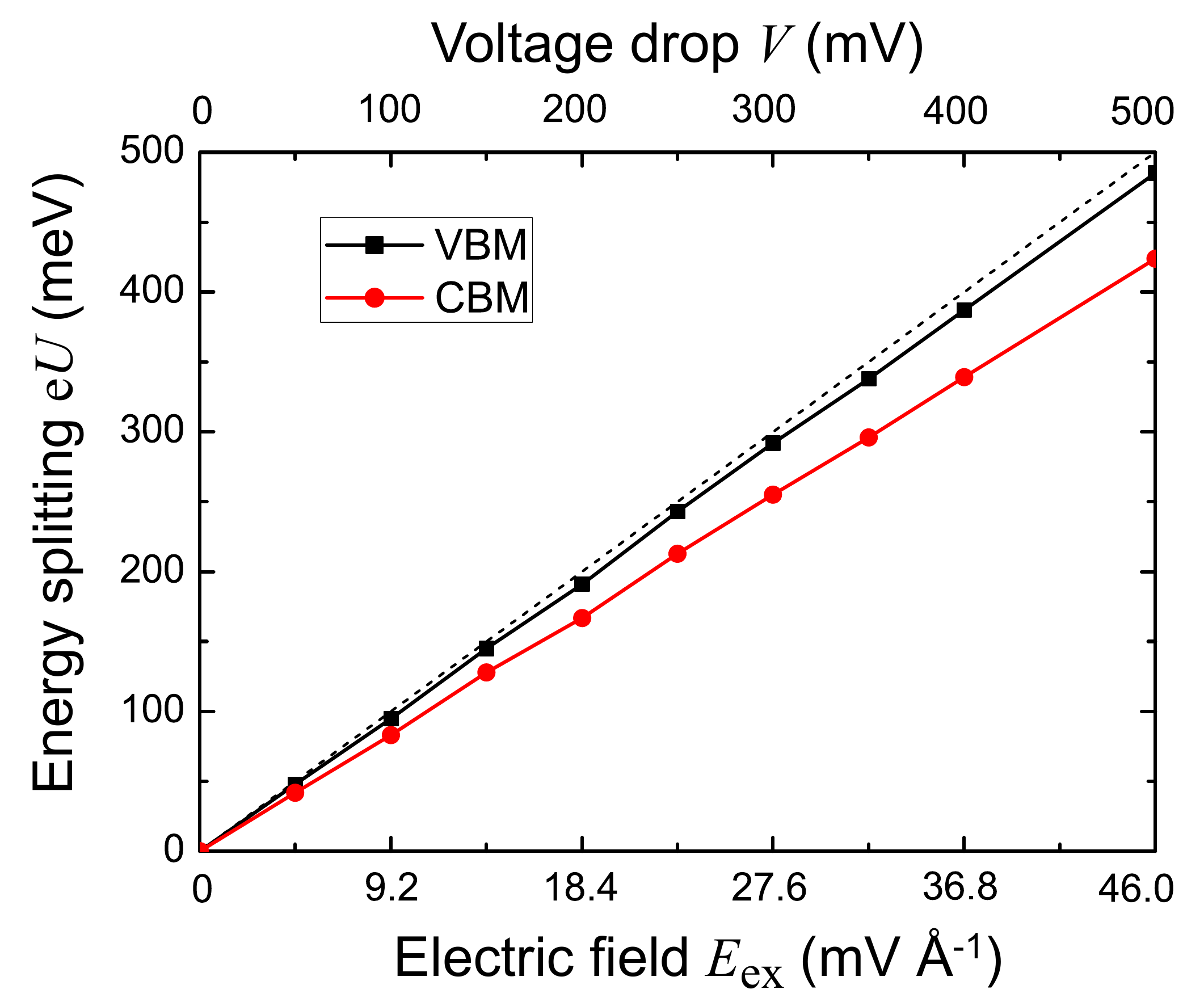}
\caption{{\bf Energy splitting vs. Electric field.} The relationship between the applied electric field $E_{\mathrm{ex}}$ (as well as the corresponding voltage drop between the two SbSe layers) and the atomic layer energy splitting e$U$. The black and the red are the energy splitting for the valence band maximum (VBM) and the conduction band minimum (CBM), respectively.}  \label{FigS12}
\end{figure}

\section{The tight-binding model}
In this section, we show the details of the tight-binding (TB) model for the bilayer SbSe checkerboard lattice. The bilayer lattice structure and three $p$ orbitals we consider in the TB model are shown in Supplementary Figure~\ref{FigS3}. There are four atoms in one unit cell, two Sb and two Se, and we denote them as $\eta_{j}$, where $\eta=$Sb,Se and $j=\pm$ refer to the upper and lower layer. The Sb (Se) atom in the upper layer has the same position in the $xy$ plane as the Se (Sb) atom in the lower layer. Thus, the lower layer can be related to the upper layer by a glide operation. The atom positions are denoted as $\mathbf{r}_{\mathbf{n}\eta_j}=\mathbf{R}_{\mathbf{n}}+\mathbf{r}_{\eta_j}$, where $\mathbf{R}_{\mathbf{n}}$ is the lattice vector and $\mathbf{r}_{\eta_j}$ is the relative position of the atoms in one unit cell. The atomic orbital wave function is  $\phi_{\alpha}(\mathbf{r}-\mathbf{r}_{\mathbf{n}\eta_j})$, where $\alpha=p_x, p_y,p_z$ for three $p$ orbitals.
The form of the TB model in the real space is
\begin{eqnarray}
  &&H_0=H_++H_-+H_{+-}\\
  &&H_{j(j=\pm)}=\sum_{\mathbf{n},\eta_j,\alpha}\epsilon_{\alpha\eta_j } c^{\dag}_{\alpha \mathbf{n}\eta_j}c_{\alpha\mathbf{n}\eta_j}+\sum_{\langle \mathbf{n}\mathbf{m}\rangle_{in},\eta_j\eta'_j,\alpha\beta}t^{\alpha\beta}_ {\mathbf{n}\eta_j,\mathbf{m}\eta'_j}c^{\dag}_{\alpha \mathbf{n}\eta_j}c_{\beta\mathbf{m}\eta'_j}\\
  &&H_{+-}=\sum_{\langle \mathbf{n}\mathbf{m}\rangle_{\pm},\eta_+\eta_-,\alpha\beta,} (r^{\alpha\beta}_{\mathbf{n}\eta_+,\mathbf{m}\eta_-} c^{\dag}_{\alpha\mathbf{n}\eta_+}c_{\beta\mathbf{m}\eta_-}+h.c.)
\end{eqnarray}
where $\langle \mathbf{n}\mathbf{m}\rangle_{in}$ refers to the hopping within the same layer, and $\langle \mathbf{n}\mathbf{m}\rangle_{\pm}$ refers to the hopping between two layers.

The basis in the momentum space can be constructed as
\begin{eqnarray}
  |\alpha\eta_j,\mathbf{k}\rangle=\frac{1}{\sqrt{N}}\sum_{\mathbf{n}}e^{i\mathbf{k} \cdot\mathbf{r}_{\mathbf{n}\eta_j}}\phi_{\alpha}(\mathbf{r}-\mathbf{r}_{\mathbf{n}\eta_j}).
\end{eqnarray}
With the transformations
\begin{eqnarray}
  &&c^{\dag}_{\alpha \mathbf{n}\eta_j}=\frac{1}{\sqrt{N}}\sum_{\mathbf{k}}e^{-i\mathbf{k} \cdot\mathbf{r}_{\mathbf{n}\eta_j}}c^{\dag}_{\alpha\eta_j}(\mathbf{k})\\
  &&c_{\beta \mathbf{m}\eta'_j}=\frac{1}{\sqrt{N}}\sum_{\mathbf{k}'}e^{i\mathbf{k}' \cdot\mathbf{r}_{\mathbf{m}\eta'_j}}c_{\beta\eta'_j}(\mathbf{k}'),
\end{eqnarray}
we can get the TB Hamiltonian in the momentum space as
\begin{eqnarray}
  &&H_0(\mathbf{k})=H_+(\mathbf{k})+H_-(\mathbf{k})+H_{+-}(\mathbf{k})\\
  &&H_{j(j=\pm)}(\mathbf{k})=\sum_{\eta_j,\alpha} \epsilon_{\alpha\eta_j} c^{\dag}_{\alpha\eta_j}(\mathbf{k})c_{\alpha\eta_j}(\mathbf{k}) +\sum_{\mathbf{\delta}_j,\eta_j\eta'_j,\alpha\beta} t^{\alpha\beta}_ {\mathbf{\delta}_j} e^{-i\mathbf{k}\cdot(\mathbf{r}_{\mathbf{n}\eta_j}-\mathbf{r}_{\mathbf{m}\eta'_j})} c^{\dag}_{\alpha\eta_j}(\mathbf{k})c_{\beta\eta'_j}(\mathbf{k})\\
  &&H_{+-}(\mathbf{k})=\sum_{\mathbf{\delta}_{\pm},\eta_+\eta_-,\alpha\beta} (r^{\alpha\beta}_{\mathbf{\delta}_{\pm}} e^{-i\mathbf{k}\cdot(\mathbf{r}_{\mathbf{n}\eta_+}-\mathbf{r}_{\mathbf{m}\eta_-})} c^{\dag}_{\alpha\eta_+}(\mathbf{k})c_{\beta\eta_-}(\mathbf{k})+h.c.)
\end{eqnarray}
where $\mathbf{\delta}_j=(\mathbf{r}_{\mathbf{n}\eta_j}-\mathbf{r}_{\mathbf{m}\eta'_j})$ with $\mathbf{m}$ and $\mathbf{n}$ in the same layer, and $\mathbf{\delta}_{\pm}=(\mathbf{r}_{\mathbf{n}\eta_+}-\mathbf{r}_{\mathbf{m}\eta_-})$ with $\mathbf{m}$ and $\mathbf{n}$ in different layers.

In the basis $\Psi_1=(|p_x,\mathrm{Sb}_+,\mathbf{k}\rangle, |p_y,\mathrm{Sb}_+,\mathbf{k}\rangle, |p_z,\mathrm{Sb}_+,\mathbf{k}\rangle, |p_x,\mathrm{Sb}_-,\mathbf{k}\rangle, |p_y,\mathrm{Sb}_-,\mathbf{k}\rangle, |p_z,\mathrm{Sb}_-\mathbf{k}\rangle,\\ |p_x,\mathrm{Se}_+,\mathbf{k}\rangle, |p_y,\mathrm{Se}_+,\mathbf{k}\rangle, |p_z,\mathrm{Se}_+\mathbf{k}\rangle, |p_x,\mathrm{Se}_-,\mathbf{k}\rangle, |p_y,\mathrm{Se}_-,\mathbf{k}\rangle,  |p_z,\mathrm{Se}_-,\mathbf{k}\rangle)^T$, the matrix form of the Hamiltonian is
\begin{eqnarray}
 H_0(\mathbf{k})=\left(
 \begin{array}{cccccccccccc}
 P_{\mathrm{Sb}}& 0 & 0 & F_1 & 0 & A_1 & J & F & 0 & M_1 & 0 & B_1 \\
 0 & Q_{\mathrm{Sb}} & 0 & 0 & F_1 & A_2 & F & J & 0 & 0 & M_2 & B_2\\
0 & 0 & L_{\mathrm{Sb}} & A_1 & A_2 & C_1 & 0 & 0 & T & B_1 & B_2 & V_1\\
 F_1 & 0 & -A_1 & P_{\mathrm{Sb}} & 0 & 0 & M_1 & 0 & -B_1 & J & F & 0\\
 0 & F_1 & -A_2 & 0 & Q_{\mathrm{Sb}} & 0 & 0 & M_2 & -B_2 & F & J & 0\\
 -A_1 & -A_2 & C_1 & 0 & 0 & L_{\mathrm{Sb}} & -B_1 & -B_2 & V_1 & 0 & 0 & T\\
 J & F & 0 & M_1 & 0 & B_1 & P_{\mathrm{Se}} &0 & 0 & F_{12} & 0 & A_{12}\\
 F & J & 0 & 0 & M_2 & B_2 & 0 & Q_{\mathrm{Se}} & 0 & 0 & F_{12} & A_{22}\\
 0 & 0 & T & B_1 & B_2 & V_1 & 0 & 0 & L_{\mathrm{Se}} & A_{12} & A_{22} & C_{12}\\
M_1 & 0 & -B_1 & J & F & 0 & F_{12} & 0 & -A_{12} & P_{\mathrm{Se}} & 0 & 0\\
 0 & M_2 & -B_2 & F & J & 0 & 0 & F_{12} & -A_{22} & 0 & Q_{\mathrm{Se}} & 0\\
-B_1 & -B_2 & V_1 & 0 & 0 & T & -A_{12} & -A_{22} & C_{12} & 0 & 0 & L_{\mathrm{Se}}\\
 \end{array}
\right)
\label{eq:H}
\end{eqnarray}
where
\begin{eqnarray}
  &&P_{\mathrm{Sb}}=\epsilon_{\mathrm{Sb}xy}+2\sigma_{2}\cos(k_x a)+2\pi_{2}\cos(k_y a), \nonumber\\
  &&P_{\mathrm{Se}}=\epsilon_{\mathrm{Se}xy}+2\sigma_{3}\cos(k_x a)+2\pi_{3}\cos(k_y a),\nonumber\\
  &&Q_{\mathrm{Sb}}=\epsilon_{\mathrm{Sb}xy}+2\pi_{2}\cos(k_x a)+2\sigma_{2}\cos(k_y a),\nonumber\\
  &&Q_{\mathrm{Se}}=\epsilon_{\mathrm{Se}xy}+2\pi_{3}\cos(k_x a)+2\sigma_{3}\cos(k_y a),\nonumber\\
  &&L_{\mathrm{Sb}}=\epsilon_{\mathrm{Se}z}+2\pi_{3}\cos(k_x a)+2\pi_{3}\cos(k_y a),\nonumber\\
  &&J=2\cos(k_x a/2)\cos(k_y a/2)(\sigma_1+\pi_1),\nonumber\\
  &&T=4\cos(k_x a/2)\cos(k_y a/2)\pi_1,\nonumber\\
  &&F=-2\sin(k_x a/2)\sin(k_y a/2)(\sigma_1-\pi_1),\nonumber\\
  &&F_1=2\cos(k_x a/2)\cos(k_y a/2)(\pi_5(1+\cos^2\gamma) +\sigma_5\sin^2\gamma),\nonumber\\
  &&F_{12}=2\cos(k_x a/2)\cos(k_y a/2)(\pi'_{5}(1+\cos^2\gamma) +\sigma'_{5}\sin^2\gamma),\nonumber\\
  &&M_1=\pi_4+2\cos(k_x a)(\sigma_{6}\sin^2\omega +\pi_{6}\cos^2\omega)+2\cos(k_y a)\pi_{6},\nonumber\\
  &&M_2=\pi_4+2\cos(k_y a)(\sigma_{6}\sin^2\omega +\pi_{6}\cos^2\omega)+2\cos(k_x a)\pi_{6},\nonumber\\
  &&A_1=-\sqrt{2}i\sin(k_x a/2)\cos(k_y a/2)(\sigma_5-\pi_5) \sin(2\gamma),\nonumber\\
   &&A_{12}=-\sqrt{2}i\sin(k_x a/2)\cos(k_y a/2)(\sigma'_5-\pi'_5) \sin(2\gamma),\nonumber\\
  &&A_2=-\sqrt{2}i\cos(k_x a/2)\sin(k_y a/2)(\sigma_5-\pi_5)\sin(2\gamma),\nonumber\\
  &&A_{22}=-\sqrt{2}i\cos(k_x a/2)\sin(k_y a/2)(\sigma'_5-\pi'_5) \sin(2\gamma),\nonumber\\
  &&B_1=-i\sin(k_x a)(\sigma_{6}-\pi_{6})\sin(2\omega),\nonumber\\
  &&B_2=-i\sin(k_y a)(\sigma_{6}-\pi_{6})\sin(2\omega),\nonumber\\
   &&C_1=4\cos(k_x a/2)\cos(k_y a/2)(\sigma_5\cos^2\gamma+\pi_5\sin^2\gamma),\nonumber\\
  &&V_1=\sigma_4+2\cos(k_x a)(\sigma_{6}\cos^2\omega+\pi_{6}\sin^2\omega)+2\cos(k_y a)(\sigma_{6}\cos^2\omega+\pi_{6}\sin^2\omega),\nonumber\\
  &&C_{12}=4\cos(k_x a/2)\cos(k_y a/2)(\sigma'_5\cos^2\gamma+\pi'_5\sin^2\gamma).
\end{eqnarray}
where $\gamma=\arctan(\sqrt{2}a/2c')$, $\omega=\arctan(a/c')$ with $a$ is the lattice constant in $x(y)$ direction and $c'$ is the distance between two layers.
$\sigma$, $\sigma'$, $\pi$, $\pi'$ and $\epsilon$ are material dependent parameters.

Then we include the atomic spin-orbit coupling term $ H_{\mathrm{SO}}=\lambda_{\mathrm{SO}}\mathbf{S}\cdot\mathbf{L} =\lambda_{\mathrm{SO}}\frac{1}{2}\bm{\sigma}\cdot\mathbf{L}$, with
\begin{eqnarray}
  &&L_x=\left(
  \begin{array}{ccc}
  0 & 0& 0\\
  0 & 0& -i \\
  0 & i & 0\\
   \end{array}
   \right);
   \qquad L_y=\left(
   \begin{array}{ccc}
     0 & 0 & i\\
     0 & 0 & 0\\
     -i & 0 & 0\\
   \end{array}
   \right);
   \qquad L_z=\left(
  \begin{array}{ccc}
  0 & -i & 0\\
  i & 0 & 0\\
  0 & 0 & 0\\
   \end{array}
   \right)
\end{eqnarray}
in the basis $(|p_x\rangle, |p_y\rangle, |p_z\rangle)^T$.
Define
\begin{eqnarray}
 && \lambda=\left(
  \begin{array}{cccc}
    \lambda_{\mathrm{Sb}} & 0 & 0 & 0\\
    0 & \lambda_{\mathrm{Sb}} & 0 & 0\\
    0 & 0 & \lambda_{\mathrm{Se}} & 0\\
    0 & 0 & 0 & \lambda_{\mathrm{Se}}\\
  \end{array}
  \right)
\end{eqnarray}
In the basis $\Psi_2=(|\uparrow\rangle, |\downarrow\rangle)^T \otimes(|p_x,\mathrm{Sb}_+,\mathbf{k}\rangle, |p_y,\mathrm{Sb}_+,\mathbf{k}\rangle, |p_z,\mathrm{Sb}_+,\mathbf{k}\rangle, |p_x,\mathrm{Sb}_-,\mathbf{k}\rangle, |p_y,\mathrm{Sb}_-,\mathbf{k}\rangle, \\|p_z,\mathrm{Sb}_-\mathbf{k}\rangle, |p_x,\mathrm{Se}_+,\mathbf{k}\rangle, |p_y,\mathrm{Se}_+,\mathbf{k}\rangle, |p_z,\mathrm{Se}_+\mathbf{k}\rangle, |p_x,\mathrm{Se}_-,\mathbf{k}\rangle, |p_y,\mathrm{Se}_-,\mathbf{k}\rangle,  |p_z,\mathrm{Se}_-,\mathbf{k}\rangle)^T$, the spin-orbit coupling term is
\begin{eqnarray}
  H_{SO}=\frac{1}{2}\big[\sigma_x\otimes\lambda\otimes L_x+\sigma_y\otimes\lambda\otimes L_y+\sigma_z\otimes\lambda\otimes L_z\big]\nonumber\\
\end{eqnarray}
where $\sigma$ denotes spin, $\lambda$ denotes the atoms and $L$ denotes the orbital angular momentum.
Now we add an asymmetric potential e$U$ on the upper layer and $-$e$U$ on the lower layer, $H_U=$e$U I_{2\times2}\otimes\tau_z\otimes I_{3\times3}$. The total Hamiltonian is given by $H=I_{2\times 2}\otimes H_0+H_{SO}+H_U$.

The energy bands calculated by the TB model are shown in Supplementary Figure~\ref{FigS4} and Figure~3(c)(d) in the main text. The parameters we use to fit the dispersions from the {\it ab initio} method are $a=4.13$, $c'=11.12$, $\epsilon_{Sbxy}=0.4$, $\epsilon_{Sexy}=-0.36$, $\epsilon_{Sbz}=0.43$, $\epsilon_{Sez}=-0.45$, $\lambda_{Sb}=0.4$, $\lambda_{Se}=0.22$, $\sigma_1=0.85$, $\pi_1=-0.09$, $\sigma_2=0.2$, $\pi_2=0.06$, $\sigma_3=-0.2$, $\pi_3=-0.02$, $\sigma_4=-1.1$, $\pi_4=-0.085$, $\sigma_5= -0.065$, $\pi_5= -0.143$, $\sigma'_{5}= 0.476$, $\pi'_5= -0.0065$, $\sigma_6=0.0026$, $\pi_6=-0.0027$.  Compare Supplementary Figure~\ref{FigS4}(a)-(c) with Figure 1b in the main text and Figure 2b,c in the main text, we find that the low energy behavior of bands calculated by the TB model are qualitatively similar to that calculated by the {\it ab initio} method with or without the asymmetric potential.
\begin{figure} [h]
\includegraphics[width=0.6\textwidth]{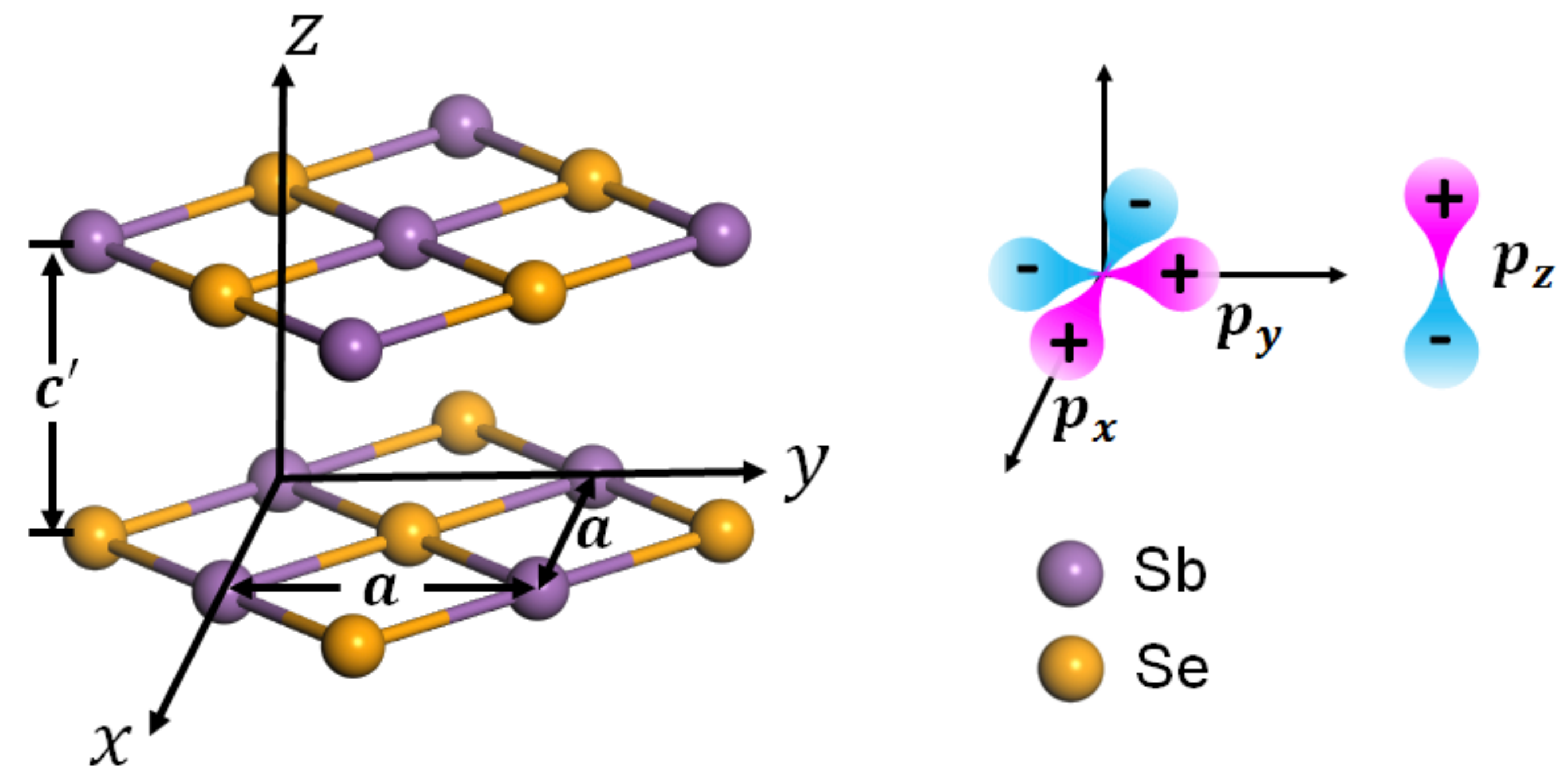}
\caption{{\bf The bilayer lattice structure used in the TB model and $p$-orbitals.} } \label{FigS3}
\end{figure}

\begin{figure} [h]
\includegraphics[width=0.6\textwidth]{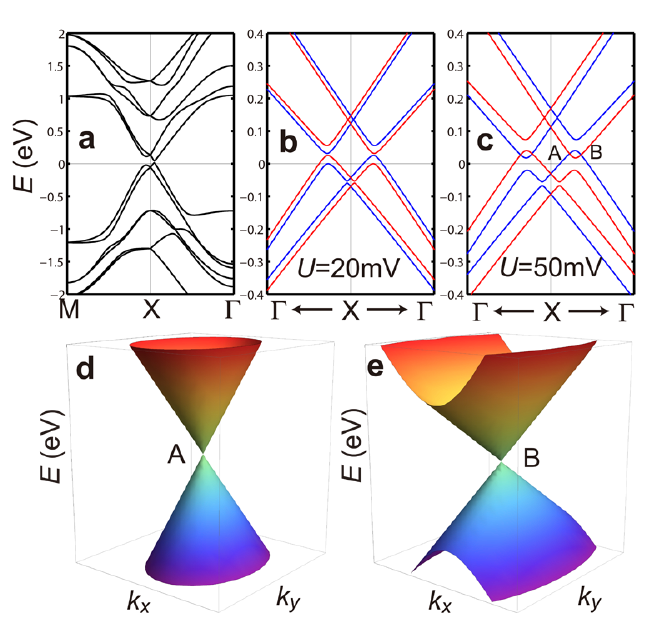}
\caption{{\bf The band structures calculated by the TB model.} (a) The bands along the high symmetry lines M--X and X--$\Gamma$ without the asymmetric potential ($U=0$). (b) and (c) The bands around the X point along the $\Gamma$--X line in the first and second BZ with asymmetric potential $U=20$ mV in (b) and $U=50$ mV in (c). The red (blue) lines refer to the bands with mirror parity $+i(-i)$ of $\hat{m}_y$. (d) and (e) The 3D Dirac cones calculated by the effective model around the A and B Dirac points in (c).} \label{FigS4}
\end{figure}

\section{Proof of the Dirac cones}
To prove that the crossings of the two bands near the band gap after we apply the asymmetric potential are indeed the Dirac cones, we derive the effective model around the crossing point from the TB model $H$. Take $U=50$ mV as an example. The band structure in this case are shown in Supplementary Figure~\ref{FigS4}(c).
We first diagonalize the Hamiltonian at the position of the crossing point to get the matrix of eigen vectors $R$ and the eigen states $|\phi_1\rangle$ and $|\phi_2\rangle$ of the two crossing bands. Perform the rotation $R$, and get $H_1=R^{\dag}HR$. Then we can project $H_1$ into the sub-space of $|\phi_1\rangle$ and $|\phi_2\rangle$ with the second order L$\ddot{o}$wdin perturbation to obtain the effective model around the crossing point. We only keep the $\mathbf{k}$ linear term here.

For the crossing point A and B in Supplementary Figure~\ref{FigS4}(c), the effective model has the form
\begin{eqnarray}
  H_{\mathrm{D},\beta(\beta=\mathrm{A},\mathrm{B})}=\varepsilon_\beta+c_{0,\beta}k_x +c_{1,\beta}k_x\hat{\sigma}_z+c_{2,\beta}k_y\hat{\sigma}_y
\end{eqnarray}
where $\varepsilon_\mathrm{A}=0.027$, $c_{0,\mathrm{A}}=0.235$, $c_{1,\mathrm{A}}=-1.825$, $c_{2,\mathrm{A}}=0.93$ and $\varepsilon_\mathrm{B}=0.03$, $c_{0,\mathrm{B}}=0.265$, $c_{1,\mathrm{B}}=-2.055$, $c_{2,\mathrm{B}}=-1.56$. The effective model around the crossing point indeed has the form of the Dirac Hamiltonian.
The band dispersion calculated by the effective Hamiltonian $H_{\mathrm{D},\beta}$ are shown in Supplementary Figure~\ref{FigS4}(d) and (e) for $\beta=\mathrm{A}$ and $\mathrm{B}$, respectively.

\section{Orbital analysis of (LaO)$_2$(SbSe$_2$)$_2$ and other (LaO)$_2$(SbSe$_2$)$_2$ class of materials}

Based on the first-principles calculations, we show the band structures of (LaO)$_2$(SbSe$_2$)$_2$ with the SbSe$_2$-layer atoms and three $p$ orbitals projections in Supplementary Figure~\ref{FigS1}. It is concluded that the states for conduction band minimum (CBM) and valence band maximum (VBM) near the X point are dominated by the $p_x$ orbitals of the Sb$_1$/Sb$^{\prime}_1$ and Se$_2$/Se$^{\prime}_2$ atoms.

We also present the band structure calculations of other materials in the (LaO)$_2$(SbSe$_2$)$_2$ family, including (LaO)$_2$(SbS$_2$)$_2$, (LaO)$_2$(SbTe$_2$)$_2$, (LaO)$_2$(BiS$_2$)$_2$\cite{tanryverdiev1995}, (LaO)$_2$(BiSe$_2$)$_2$, (LaO)$_2$(BiTe$_2$)$_2$ and (SrF)$_2$(SbSe$_2$)$_2$\cite{kabbour2006}. The calculation of (SrF)$_2$(SbSe$_2$)$_2$ reveals that the (LaO)$_2^{2+}$ layer can be replaced by the (SrF)$_2^{2+}$ layer. Experimental lattice constants are used when available, and for the compounds that have not been fabricated, fully relaxed calculations are adopted to obtain their optimal lattice constants. Since the energy dispersion along the $k_z$ direction is almost flat, we only show their band structures along the M--X--$\Gamma$ lines in Supplementary Figure~\ref{FigS2}. All materials have the similar electronic properties of CBM, but different locations of VBM. For (LaO)$_2$(Sb$X_2$)$_2$ ($X$ = S, Se or Te), both CBM and VBM are located in the vicinity of X point. Therefore, all these materials can be used to realize electrically tunable Dirac cones. For (LaO)$_2$(Bi$X_2$)$_2$ ($X$ = S, Se or Te), there are indirect gaps between CBM near the X point and VBM at the $\Gamma$ point. Thus, for these materials, the system will become semi-metal with electron pocket near the X point and hole pocket near the $\Gamma$ point when applying gate voltages. All states near the bandgap are dominated by the $p$ orbitals in ($MX_2$) layers.
\begin{figure} [h]
\includegraphics[width=0.6\textwidth]{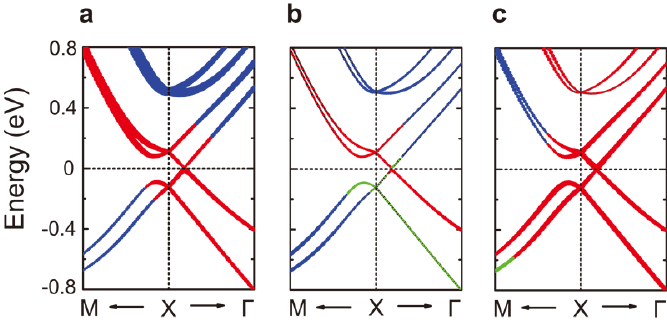}
\caption{{\bf Atomic orbital projection of the band structure.} Energy dispersion of (LaO)$_2$(SbSe$_2$)$_2$ with the Sb$_1$/Sb$^{\prime}_1$, Se$_1$/Se$^{\prime}_1$ and Se$_2$/Se$^{\prime}_2$ atom projections for (a), (b) and (c), respectively. The red, blue and green circles represent the weights of the $p_x$, $p_y$ and $p_z$ orbital characters, respectively.} \label{FigS1}
\end{figure}
\section{The effective model around X point}
In this section, we derive the effective model for the conduction bands and valence bands near the band gap by the perturbation method. The start point is the TB model we obtained in Supplementary Note 2. As is shown in Supplementary Figure~\ref{FigS1}, the bands near the band gap are dominated by the $p_x$ orbital of Sb and Se in the checkerboard lattice.

We first expand the elements of the TB Hamiltonian around X$=(\pi/a,0)$ point to the linear order in $k_x$ and $k_y$ as
\begin{eqnarray}
  &&P_{\mathrm{Sb}}\approx P'_{\mathrm{Sb}} =\epsilon_{\mathrm{Sb}xy}-2\sigma_{2}+2\pi_{2}, \nonumber\\
  &&P_{\mathrm{Se}}\approx P'_{\mathrm{Se}} =\epsilon_{\mathrm{Se}xy}-2\sigma_{3}+2\pi_{3},\nonumber\\
  &&Q_{\mathrm{Sb}}\approx Q'_{\mathrm{Sb}} =\epsilon_{\mathrm{Sb}xy}-2\pi_{2}+2\sigma_{2},\nonumber\\
  &&Q_{\mathrm{Se}}\approx Q'_{\mathrm{Se}} =\epsilon_{\mathrm{Se}xy}-2\pi_{3}+2\sigma_{3},\nonumber\\
  &&L_{\mathrm{Sb}}\approx L'_{\mathrm{Sb}}=\epsilon_{\mathrm{Sb}z},\nonumber\\
  &&L_{\mathrm{Se}}\approx L'_{\mathrm{Se}}=\epsilon_{\mathrm{Se}z},\nonumber\\
  &&J\approx J'k_x=-a(\sigma_1+\pi_1)k_x,\nonumber\\
  &&T\approx T'k_x =-2a\pi_1k_x,\nonumber\\
  &&F\approx F'k_y=-a(\sigma_1-\pi_1)k_y,\nonumber\\
  &&F_1\approx F'_1k_x=-a(\pi_5(1+\cos^2\gamma) +\sigma_5\sin^2\gamma)k_x,\nonumber\\
  &&F_{12}\approx F'_{12}k_x=-a(\pi'_{5}(1+\cos^2\gamma) +\sigma'_{5}\sin^2\gamma)k_x,\nonumber\\
    &&M_1\approx M'_1=\pi_4-2(\sigma_{6}\sin^2\omega +\pi_{6}\cos^2\omega)+2\pi_{6},\nonumber\\
  &&M_2\approx M'_2=\pi_4+2(\sigma_{6}\sin^2\omega +\pi_{6}\cos^2\omega)-2\pi_{6},\nonumber\\
  &&A_1\approx iA'_1=-i\sqrt{2}(\sigma_5-\pi_5) \sin(2\gamma),\nonumber\\
   &&A_{12}\approx iA'_{12}=-i\sqrt{2}(\sigma'_5-\pi'_5) \sin(2\gamma),\nonumber\\
  &&A_2\approx 0,\nonumber\\
  &&A_{22}\approx 0,\nonumber\\
  &&B_1\approx iB'_1 k_x=ia(\sigma_{6}-\pi_{6})\sin(2\omega)k_x,\nonumber\\
  &&B_2\approx iB'_2 k_y=-ia(\sigma_{6}-\pi_{6})\sin(2\omega)k_y,\nonumber\\
   &&C_1\approx C'_1 k_x=-2a(\sigma_5\cos^2\gamma+\pi_5\sin^2\gamma)k_x ,\nonumber\\
  &&V_1\approx V'_1=\sigma_4,\nonumber\\
  &&C_{12}\approx C'_{12}k_x =-2a(\sigma'_5\cos^2\gamma+\pi'_5\sin^2\gamma) k_x.
\end{eqnarray}
where $k_x$ and $k_y$ are measured relative to the X point.

Since there is a strong hybridization between the $p_x$ orbital of Sb and Se for both the conduction band and valence band, we need to find the eigenstates at X point and use them as the basis of the effective model.
We use the L$\ddot{o}$wdin perturbation method to project the $p_y$ and $p_z$ orbital onto the $p_x$ orbital. Then we get a $8\times 8$ matrix for the lowest four conduction bands and the highest four valence bands. The eigenstates of this matrix at X point can be found easily. Thus, we obtain the zeroth order basis at X point as
\begin{eqnarray}
  &&|p_x,\uparrow_z,\mathrm{c}_+\rangle=g_1|p_x,\uparrow_z,\mathrm{Sb}_-\rangle +g_2|p_x,\uparrow_z,\mathrm{Se}_+\rangle\nonumber\\
  &&|p_x,\uparrow_z,\mathrm{c}_-\rangle=g_1|p_x,\uparrow_z,\mathrm{Sb}_+\rangle +g_2|p_x,\uparrow_z,\mathrm{Se}_-\rangle\nonumber\\
  &&|p_x,\downarrow_z,\mathrm{c}_+\rangle=g_1|p_x,\downarrow_z,\mathrm{Sb}_-\rangle +g_2|p_x,\downarrow_z,\mathrm{Se}_+\rangle\nonumber\\
  &&|p_x,\downarrow_z,\mathrm{c}_-\rangle=g_1|p_x,\downarrow_z,\mathrm{Sb}_+\rangle +g_2|p_x,\downarrow_z,\mathrm{Se}_-\rangle\nonumber\\
  &&|p_x,\uparrow_z,\mathrm{v}_+\rangle=g_2|p_x,\uparrow_z,\mathrm{Sb}_+\rangle -g_1|p_x,\uparrow_z,\mathrm{Se}_-\rangle\nonumber\\
  &&|p_x,\uparrow_z,\mathrm{v}_-\rangle=g_2|p_x,\uparrow_z,\mathrm{Sb}_-\rangle -g_1|p_x,\uparrow_z,\mathrm{Se}_+\rangle\nonumber\\
  &&|p_x,\downarrow_z,\mathrm{v}_+\rangle=g_2|p_x,\downarrow_z,\mathrm{Sb}_+\rangle -g_1|p_x,\downarrow_z,\mathrm{Se}_-\rangle\nonumber\\
  &&|p_x,\downarrow_z,\mathrm{v}_-\rangle=g_2|p_x,\downarrow_z,\mathrm{Sb}_-\rangle -g_1|p_x,\downarrow_z,\mathrm{Se}_+\rangle
\end{eqnarray}
Here $g_2/g_1=[-P_{\mathrm{Sb}1}+P_{\mathrm{Se}1}+ \sqrt{4M'_1+(P_{\mathrm{Sb}1}-P_{\mathrm{Se}1})^2}]/(2M'_1)$, where
\begin{eqnarray}
  &&P_{\mathrm{Sb}1}=P'_{\mathrm{Sb}}+\frac{(A'_1)^2}{P'_{\mathrm{Sb}}-L'_{\mathrm{Sb}}} +\frac{1}{4}\left(\frac{1}{P'_{\mathrm{Sb}}-L'_{\mathrm{Sb}}} +\frac{1}{P'_{\mathrm{Sb}}-Q'_{\mathrm{Sb}}}\right)\lambda_{\mathrm{Sb}}^2\nonumber\\
  &&P_{\mathrm{Se}1}=P'_{\mathrm{Se}}+\frac{(A'_{12})^2}{P'_{\mathrm{Se}}-L'_{\mathrm{Se}}} +\frac{1}{4}\left(\frac{1}{P'_{\mathrm{Se}}-L'_{\mathrm{Se}}} +\frac{1}{P'_{\mathrm{Se}}-Q'_{\mathrm{Se}}}\right)\lambda_{\mathrm{Sb}}^2
\end{eqnarray}
And $g_1\approx-0.21$, $g_2\approx0.98$ with the TB parameters we use.

Using the L$\ddot{o}$wdin perturbation method again, we obtain the $4\times 4$ effective model for the lowest conduction bands (c) (the highest valence bands (v)) as
\begin{eqnarray} H_{\xi=\mathrm{c},\mathrm{v}}(\mathbf{k})
&=&\left(\begin{array}{cccc}
\epsilon_{\xi}+V & f_{1\xi}k_x & -if_{2\xi} k_x-f_{3\xi}k_y &0\\
f_{1\xi}k_x & \epsilon_{\xi}-V & 0 & if_{2\xi}k_x+f_{3\xi}k_y\\
if_{2\xi}k_x-f_{3\xi}k_y &0 & \epsilon_{\xi}+V &f_{1\xi}k_x\\
0 & -if_{2\xi}k_x+f_{3\xi}k_y &f_{1\xi}k_x &\epsilon_{\xi}-V\\
\end{array}
\right)\nonumber\\
&=&\epsilon_{\xi}(k_x,k_y)+f_{1\xi}k_x\hat{\tau}_x +f_{2\xi}k_x\hat{\sigma}_y\hat{\tau}_z-f_{3\xi}k_y\hat{\sigma}_x \hat{\tau}_z+U\hat{\tau}_z
\label{eq:Heff}
\end{eqnarray}
up to the second order in $\mathbf{k}$ with the zeroth order basis $|p_x,\sigma,\xi_\mu\rangle$, where $\sigma=\uparrow_z,\downarrow_z$, $\xi=\mathrm{c},\mathrm{v}$ and $\mu=\pm$. $k_x$ and $k_y$ are measured relative to the X point. The diagonal term $\epsilon_{\xi}(k_x,k_y)=d_{0\xi}+d_{1\xi}k_x^2+d_{2\xi}k_y^2$ is not important. The second term mainly comes from the second order perturbation. The expression of the largest term in $f_{1c}$ is
\begin{eqnarray}
  &&f^{(2)}_{1\mathrm{c}}\approx-\frac{2((F'_1-F'_{12})g_1g_2+ (-g_1^2+g_2^2)J')((g_1^2-g_2^2)M'_1+g_1g_2 (-P'_{\mathrm{Sb}}+P'_{\mathrm{Se}}))}{4g_1g_2M_1+(g_1^2-g_2^2) (P_{\mathrm{Sb}}-P_{\mathrm{Se}})}
\end{eqnarray}
Since $|g_2|>|g_1|$, the most important hopping is between the $p_x$ orbital of Sb and Se in the same layer ($J'$) and in different layers ($M'_1$). The third and fourth term mainly come from the third order perturbation. The expression of the largest term in $f_{2c}$ is
\begin{eqnarray}
 &&f^{(3)}_{2\mathrm{c}}\approx\frac{A'_{12}C'_{12}g_2^2 \lambda_{\mathrm{Se}}}{(-L'_{\mathrm{Se}}+2g_1g_2M'_1+g_1^2P'_{\mathrm{Sb}} +g_2^2P'_{\mathrm{Se}})^2}
\end{eqnarray}
Thus, $f_{2\mathrm{c}}$ are dominated by the hopping between the $p_x$ and $p_z$ orbital of Se in different layers ($A'_{12}$), the hopping between $p_z$ orbital of Se in different layers ($C'_{12}$) and the spin-orbit coupling effect of Se ($\lambda_{\mathrm{Se}}$).

The largest term in $f_{3\mathrm{c}}$ is
\begin{eqnarray}
  f^{(3)}_{3\mathrm{c}}&\approx& \frac{g_1g_2A'_1F'\lambda_{\mathrm{Sb}}} {(L'_{\mathrm{Sb}}-2g_1g_2M'_1-g_1^2P'_{\mathrm{Sb}}-g_2^2P'_{\mathrm{Se}}) (-Q'_{\mathrm{Sb}}+2g_1g_2M'_1+g_1^2P'_{\mathrm{Sb}}+g_2^2P'_{\mathrm{Se}})} \nonumber\\&&-\frac{g_1g_2A'_{12}F'\lambda_{\mathrm{Se}}} {(L'_{\mathrm{Se}}-2g_1g_2M'_1-g_1^2P'_{\mathrm{Sb}}-g_2^2P'_{\mathrm{Se}}) (-Q'_{\mathrm{Se}}+2g_1g_2M'_1+g_1^2P'_{\mathrm{Sb}}+g_2^2P'_{\mathrm{Se}})}
\end{eqnarray}
The coefficient $f_{3\mathrm{c}}$ are dominated by the hopping between $p_x$ and $p_z$ orbital of Sb (Se) in different layers ($A'_1$ ($A'_{12}$)), the hopping between $p_x$ and $p_y$ orbital of nearest Sb and Se in the same layer ($F$) and the spin-orbital coupling effect ($\lambda_{Sb}$, $\lambda_{\mathrm{Se}}$). The expressions of the coefficients of the valence bands have similar forms, so we will not show it here. The values of $f_{i\xi}$ and $d_{0\xi}$ are
\begin{eqnarray}
  &&f_{1\mathrm{c}}\approx-10.39,~ f_{2\mathrm{c}}\approx0.95,~ f_{3\mathrm{c}}\approx-0.088,~ d_{0\mathrm{c}}\approx0.48, ~ d_{1\mathrm{c}}\approx-883, ~ d_{2\mathrm{c}}\approx-70\nonumber\\
  &&f_{1\mathrm{v}}\approx12.86,~ f_{2\mathrm{v}}\approx-0.26,~ f_{3\mathrm{v}}\approx-0.057,~ d_{0\mathrm{v}}\approx-0.36, ~ d_{1\mathrm{v}}\approx883, ~ d_{2\mathrm{v}}\approx 66
\end{eqnarray}

The coefficients $f_{2\xi}$ and $f_{3\xi}$ determines the spin texture of the bands, and we find that for the conduction bands $f_{2\mathrm{c}}$ and $f_{3\mathrm{c}}$ have opposite sign, while for the valence bands $f_{2\mathrm{v}}$ and $f_{3\mathrm{v}}$ have the same sign. This will lead to different type of spin textures for the conduction and valence bands as shown in Supplementary Note 7.

\section{Symmetry analysis}
The effective model we get in Supplementary Note 5 through the perturbation method can also be obtained directly from the symmetry analysis based on the theory of invariants. At X point the little group includes the mirror reflection symmetry in $y$ direction $\hat{m}_y$, space inversion symmetry $\hat{I}$, and glide symmetry $\hat{g}_z$. We also have time-reversal symmetry $\hat{T}$ at this point.

First, we need to find the operation of these symmetry operators on the basis $|p_x,\sigma,\eta_{\mu},\mathbf{k}\rangle$ where $\eta=$Sb,Se. Choosing the origin of the coordinate on one Sb in the upper layer. The expressions of the basis are
\begin{eqnarray}
  &&|p_x,\sigma,\mathrm{Sb}_+,\mathbf{k}\rangle=\frac{1}{\sqrt{N}}\sum_{\mathbf{R}} e^{i\mathbf{k}\cdot\mathbf{R}} \phi(\mathbf{r}-\mathbf{R})|\sigma\rangle\\
  &&|p_x,\sigma,\mathrm{Sb}_-,\mathbf{k}\rangle=\frac{1}{\sqrt{N}}\sum_{\mathbf{R}} e^{i\mathbf{k}\cdot(\mathbf{R}+\mathbf{\tau})} \phi(\mathbf{r}-\mathbf{R}-\mathbf{\tau})|\sigma\rangle
\end{eqnarray}
where $\mathbf{\tau}=(a/2,a/2)$.

The operation of the symmetry operator $\hat{h}$ on a general wave function has the form
\begin{eqnarray}
  \hat{h}\psi=\mathcal{D}(h)\psi =\mathcal{D}_{1/2}(h)\psi(\hat{h}^{-1}\mathbf{r})
\end{eqnarray}
where $\mathcal{D}_{1/2}(h)$ is the transformation of the spinor.

Take the mirror reflection symmetry $\hat{m}_y$ as an example,
\begin{eqnarray}
  &&\hat{m}_y|p_x,\uparrow_z,\mathrm{Sb}_+,\mathbf{k}\rangle=\hat{m}_y \frac{1}{\sqrt{N}}\sum_{\mathbf{R}} e^{i\mathbf{k}\cdot\mathbf{R}} \phi(\mathbf{r}-\mathbf{R})|\uparrow_z\rangle =\frac{1}{\sqrt{N}}\sum_{\mathbf{R}} e^{i\mathbf{k}\cdot\mathbf{R}} \phi(\hat{m}_y^{-1}\mathbf{r}-\mathbf{R}) i\sigma_y|\uparrow_z\rangle\nonumber\\
  &&=-\frac{1}{\sqrt{N}}\sum_{\mathbf{R}} e^{i\hat{m}_y\mathbf{k}\cdot\hat{m}_y\mathbf{R}} \phi(\mathbf{r}-\hat{m}_y\mathbf{R})|\downarrow_z\rangle
  =-|p_x,\downarrow_z,\mathrm{Sb}_+,\hat{m}_y\mathbf{k}\rangle
\end{eqnarray}
At X point, we have $\hat{m}_y\mathbf{X}=\mathbf{X}$, thus
\begin{eqnarray}
  \hat{m}_y|p_x,\uparrow_z,\mathrm{Sb}_+,\mathbf{X}\rangle= -|p_x,\downarrow_z,\mathrm{Sb}_+,\mathbf{X}\rangle.
\end{eqnarray}
In below, we list the operation of the symmetry operators on the basis functions. For $\hat{m}_y$,
\begin{eqnarray}
  &&\hat{m}_y|p_x,\uparrow_z,Sb_{+(-)},\mathbf{X}\rangle= -|p_x,\downarrow_z,Sb_{+(-)},\mathbf{X}\rangle\nonumber\\
  &&\hat{m}_y|p_x,\downarrow_z,Sb_{+(-)},\mathbf{X}\rangle= |p_x,\uparrow_z,Sb_{+(-)},\mathbf{X}\rangle,
\end{eqnarray}
For $\hat{I}$,
\begin{eqnarray}
  &&\hat{I}|p_x,\sigma,Sb_{+},\mathbf{X}\rangle =i|p_x,\sigma,\mathrm{Sb}_-,\mathbf{X}\rangle\nonumber\\
  &&\hat{I}|p_x,\sigma,Sb_{-},\mathbf{X}\rangle =-i|p_x,\sigma,\mathrm{Sb}_+,\mathbf{X}\rangle,
\end{eqnarray}
For $\hat{g}_z$,
\begin{eqnarray}
  &&\hat{g}_z|p_x,\uparrow_z,Sb_{+(-)},\mathbf{X}\rangle= |p_x,\uparrow_z,Sb_{-(+)},\mathbf{X}\rangle\nonumber\\
  &&\hat{g}_z|p_x,\downarrow_z,Sb_{+(-)},\mathbf{X}\rangle= -|p_x,\downarrow_z,Sb_{-(+)},\mathbf{X}\rangle,
\end{eqnarray}
For $\hat{T}$,
\begin{eqnarray}
  &&\hat{T}|p_x,\uparrow_z(\downarrow_z),\mathrm{Sb}_+,\mathbf{X}\rangle= +(-)|p_x,\downarrow_z(\uparrow_z),\mathrm{Sb}_+,\mathbf{X}\rangle\nonumber\\
  &&\hat{T}|p_x,\uparrow_z(\downarrow_z),\mathrm{Sb}_-,\mathbf{X}\rangle= -(+)|p_x,\downarrow_z(\uparrow_z),\mathrm{Sb}_-,\mathbf{X}\rangle.
\end{eqnarray}
Also for any symmetry operator $\hat{h}$ in the little group of point X, its operations on the basis with $|Se_{+(-)}\rangle$ are the same with that on the basis with $|Sb_{-(+)}\rangle$, since Se$_{+(-)}$ occupies the same position with Sb$_{-(+)}$ in the $xy$-plane.

In the basis of the conduction and valence bands $|p_x,\sigma,\xi_{\mu},\mathbf{X}\rangle$, where $\xi=c,v$, the representation of the symmetry operations are
\begin{eqnarray}
 &&m_{y,c(v)}=i\hat{\sigma}_y;\qquad I_{c(v)}=-(+)\hat{\tau}_y;\qquad g_{z,c(v)}=\hat{\sigma}_z\hat{\tau}_x;\qquad T_{c(v)}=+(-)i\hat{\sigma}_y\hat{\tau}_zK
\end{eqnarray}
We can obtain the anti-commutation relation $\{\hat{m}_y,\hat{g}_z\}=0$ and $\{\hat{I},\hat{g}_z\}=0$ easily from the anti-commutation relation of the Pauli matrices.

Based on the theory of invariants, the effective Hamiltonian must be invariant under all symmetry operations of the system. Any $4\times 4$ matrix of $\mathbf{k}$ can be constructed by the linear combination of the term $\hat{\sigma}_i\hat{\tau}_j$, where $i,j=0,x,y,z$, with $\mathbf{k}$ dependent expansion coefficients. So let's check the properties of these matrices and the polynomials of $\mathbf{k}$ up to the second order under the symmetry operations.
Take $\hat{m}_y$ and $\hat{\sigma}_x\hat{\tau}_z$ as an example.
\begin{eqnarray}
\hat{m}_y(\hat{\sigma}_x\hat{\tau}_z)\hat{m}_y^{-1} =i\hat{\sigma}_y(\hat{\sigma}_x\hat{\tau}_z) (-i\hat{\sigma}_y)=-i\hat{\sigma}_x\hat{\sigma}_y \hat{\tau}_z(-i\hat{\sigma}_y)=-\hat{\sigma}_x\hat{\tau}_z \hat{\sigma}_y^2=-\hat{\sigma}_x\hat{\tau}_z
\end{eqnarray}
Thus, $\hat{m}_y$ changes $\hat{\sigma}_x\hat{\tau}_z$ to $-\hat{\sigma}_x\hat{\tau}_z$. Also we know that $\hat{m}_y$ changes $k_y$ to $-k_y$. Then the combination of them $k_y\hat{\sigma}_x\hat{\tau}_z$ will not change under $\hat{m}_y$. Following the same logic, from the list in Supplementary Table 1 we can find that the possible terms in the Hamiltonian are $(c,k_x^2,k_y^2)\hat{\sigma}_0\hat{\tau}_0$, $k_x\hat{\sigma}_0\hat{\tau}_x$, $k_x\hat{\sigma}_y\hat{\tau}_z$, and $k_y\hat{\sigma}_x\hat{\tau}_z$. Therefore, we can get the effective model with the same form in Eq.~\ref{eq:Heff}.

Now we verify the symmetry protection of the Dirac cones directly from the TB model. The Dirac nodes lie on the $\Gamma$--X line, and we have the mirror symmetry $\hat{m}_y$ on this line. We can constructed the new basis $|\alpha,\sigma',\eta_{\mu},\mathbf{k}\rangle$ where $\sigma'=\uparrow_y,\downarrow_y$ with the spin quantized in the $y$ direction as
\begin{eqnarray}
  |\alpha,\uparrow_y(\downarrow_y),\eta_{\mu}\rangle= \frac{1}{\sqrt{2}}(|\alpha,\uparrow_z,\eta_{\mu}\rangle +(-)i|\alpha,\downarrow_z,\eta_{\mu}\rangle)
\end{eqnarray}
Then we have
\begin{eqnarray}
  \hat{m}_y|\alpha,\uparrow_y(\downarrow_y), \eta_{\mu}\rangle=+(-)i(-1)^{\delta_{\alpha,p_y}} |\alpha,\uparrow_y(\downarrow_y), \eta_{\mu}\rangle
\end{eqnarray}
Therefore in the basis $\Psi_3=(|p_x,\uparrow_y,\mathrm{Sb}_+\rangle, |p_y,\downarrow_y,\mathrm{Sb}_+\rangle,|p_z,\uparrow_y,\mathrm{Sb}_+\rangle, |p_x,\uparrow_y,\mathrm{Sb}_-\rangle, |p_y,\downarrow_y,\mathrm{Sb}_-\rangle,\\|p_z,\uparrow_y,\mathrm{Sb}_-\rangle, |p_x,\uparrow_y,\mathrm{Se}_+\rangle, |p_y,\downarrow_y,\mathrm{Se}_+\rangle,|p_z,\uparrow_y,\mathrm{Se}_+\rangle, |p_x,\uparrow_y,\mathrm{Se}_-\rangle, |p_y,\downarrow_y,\mathrm{Se}_-\rangle,|p_z,\uparrow_y,\mathrm{Se}_-\rangle,\\ |p_x,\downarrow_y,\mathrm{Sb}_+\rangle, |p_y,\uparrow_y,\mathrm{Sb}_+\rangle,|p_z,\downarrow_y,\mathrm{Sb}_+\rangle, |p_x,\downarrow_y,\mathrm{Sb}_-\rangle, |p_y,\uparrow_y,\mathrm{Sb}_-\rangle,|p_z,\downarrow_y,\mathrm{Sb}_-\rangle, |p_x,\downarrow_y,\mathrm{Se}_+\rangle,\\ |p_y,\uparrow_y,\mathrm{Se}_+\rangle,|p_z,\downarrow_y,\mathrm{Se}_+\rangle, |p_x,\downarrow_y,\mathrm{Se}_-\rangle, |p_y,\uparrow_y,\mathrm{Se}_-\rangle,|p_z,\downarrow_y,\mathrm{Se}_-\rangle )^T$, the matrix of the TB Hamiltonian will be block diagonalized into two blocks on the $\Gamma$--X line. The first (second) block corresponds to the mirror parity $+i(-i)$ of $\hat{m}_y$. If a crossing is between two bands which come from different blocks, it is protected by the mirror symmetry and can not be gapped if this symmetry is preserved. In Supplementary Figure~\ref{FigS4} (b) and (c), the red (blue) bands correspond to the block with mirror parity $+i(-i)$. Without the asymmetric potential the bands come from the two blocks form degenerate pairs. When we introduce the asymmetric potential, the double degeneracy will split, but the mirror symmetry still exists. The lowest conduction band in the first BZ has mirror parity $-i$, while the highest valence band has mirror parity $+i$, so the crossings between them are protected by the mirror symmetry.
\begin{table} [h]
\renewcommand{\thetable}{\arabic{table}}
\centering
\begin{tabular}{c|c|cccc||c|c|cccc}
    \hline\hline
      &  & $T$ & $m_y$ & $g_z$ & $I$ & & & $T$ & $m_y$ & $g_z$ & $I$\\
    \hline
    $\hat{\sigma}_0\hat{\tau}_0$ &  $c,k_x^2,k_y^2$ &  + &  + &  + &  +& $\hat{\sigma}_y\hat{\tau}_x$ & $U$ & + & +&- &- \\
    $\hat{\sigma}_x\hat{\tau}_0$ &  N &  - &  - &  - &  +& $\hat{\sigma}_z\hat{\tau}_x$ & N & + & -&+ &- \\
    $\hat{\sigma}_y\hat{\tau}_0$ &  N &  - &  + &  - &  +& $\hat{\sigma}_x\hat{\tau}_y$ & N& -& -&+ &+ \\
    $\hat{\sigma}_z\hat{\tau}_0$ &  N &  - &  - &  + &  +& $\hat{\sigma}_y\hat{\tau}_y$ & N&- &+ &+ &+ \\
    $\hat{\sigma}_0\hat{\tau}_x$ &  $k_x$ &  - &  + &  + &  -& $\hat{\sigma}_z\hat{\tau}_y$ &N&- &- &- &+\\
    $\hat{\sigma}_0\hat{\tau}_y$ &  N &  + &  + &  - &  +& $\hat{\sigma}_x\hat{\tau}_z$ & $k_y$ &- & -&+ &-\\
    $\hat{\sigma}_0\hat{\tau}_z$ &  $U$ &  + &  + &  - &  -& $\hat{\sigma}_y\hat{\tau}_z$ & $k_x$&- &+&+ &-\\
    $\hat{\sigma}_x\hat{\tau}_x$ &  N &  + &  - &  - &  -& $\hat{\sigma}_z\hat{\tau}_z$ &N & -& -&- &-\\
    \hline
  \end{tabular}
\caption{{\bf Symmetry analysis.} Transformation properties of the $\hat{\sigma}_i\hat{\tau}_j$ matrices and the polynomials of $\mathbf{k}$ up to the second order under the symmetry operations.  $N$ in the table means there is no polynomials of $\mathbf{k}$ up to the second order satisfies the corresponding transformation rule. }\label{Stab:1}
\end{table}
\section{Spin polarization}
In this part, we derive the analytical expression of the layer dependent spin polarization.

Starting from the effective model (we omit $\xi$ for simplicity),
\begin{eqnarray}
  H_{eff}=\epsilon+f_1 k_x \hat{\tau}_x+(f_2 k_x \hat{\sigma}_y -f_3 k_y \hat{\sigma}_x)\hat{\tau}_z
\end{eqnarray}
we get two double degenerate levels $E_{\pm}=\epsilon \pm\sqrt{(f_1^2+f_2^2)k_x^2+f_3^2k_y^2}$.
For $E_+$, the two degenerate states are
\begin{eqnarray}
  &&\Psi_{1,+}=\frac{\sqrt{f_2^2k_x^2+f_3^2k_y^2}} {\sqrt{2}d}(\frac{if_1k_x}{f_2k_x+if_3k_y}, \frac{id}{f_2k_x+if_3k_y},0,1)\nonumber\\
  &&\Psi_{2,+}=\frac{\sqrt{f_2^2k_x^2+f_3^2k_y^2}} {\sqrt{2}d}(\frac{-id}{f_2k_x+if_3k_y}, \frac{-if_1k_x}{f_2k_x+if_3k_y} k_x,1,0)
\end{eqnarray}
where $d=\sqrt{(f_1^2+f_2^2)k_x^2+f_3^2k_y^2}$.

The layer dependent spin operator is defined as $S_{i,\mu}=\frac{\hbar}{2}\hat{\sigma}_i\frac{\hat{\tau}_z+ \mu}{2}$. For the two degenerate bands $\Psi_{1,+}$ and $\Psi_{2,+}$, we have
\begin{eqnarray}
  &&\langle \mathbf{S}_+^+\rangle=\sum_{\alpha=1,2}\langle \Psi_{\alpha,+}|\mathbf{S}_+|\Psi_{\alpha,+}\rangle= \frac{\hbar}{2d}(-f_3k_y\hat{x}+f_2k_x\hat{y})\nonumber\\
  &&\langle \mathbf{S}_-^+\rangle=\sum_{\alpha=1,2}\langle \Psi_{\alpha,+}|\mathbf{S}_-|\Psi_{\alpha,+}\rangle= \frac{\hbar}{2d}(f_3k_y\hat{x}-f_2k_x\hat{y})
\end{eqnarray}

Therefore, the spin polarization for the upper and lower layer are opposite to each other for $E_+$, that is $\langle \mathbf{S}_+^+\rangle=-\langle \mathbf{S}_-^+\rangle$.

For $E_-$, the two degenerate states are
\begin{eqnarray}
  &&\Psi_{1,-}=\frac{\sqrt{f_2^2k_x^2+f_3^2k_y^2}} {\sqrt{2}d}(\frac{if_1k_x}{f_2k_x+if_3k_y}, \frac{-id}{f_2k_x+if_3k_y},0,1)\nonumber\\
  &&\Psi_{2,-}=\frac{\sqrt{f_2^2k_x^2+f_3^2k_y^2}} {\sqrt{2}d}(\frac{id}{f_2k_x+if_3k_y}, \frac{-if_1k_x}{f_2k_x+if_3k_y} k_x,1,0)
\end{eqnarray}
For the two degenerate bands $\Psi_{1,-}$ and $\Psi_{2,-}$, we have
\begin{eqnarray}
  &&\langle \mathbf{S}_+^-\rangle=\sum_{\alpha=1,2}\langle \Psi_{\alpha,-}|\mathbf{S}_+|\Psi_{\alpha,-}\rangle= \frac{\hbar}{2d}(f_3k_y\hat{x}-f_2k_x\hat{y})\nonumber\\
  &&\langle \mathbf{S}_-^-\rangle=\sum_{\alpha=1,2}\langle \Psi_{\alpha,-}|\mathbf{S}_-|\Psi_{\alpha,-}\rangle= \frac{\hbar}{2d}(-f_3k_y\hat{x}+f_2k_x\hat{y})
\end{eqnarray}
 Similar to the case of $E_+$, we get $\langle \mathbf{S}_+^-\rangle=-\langle \mathbf{S}_-^-\rangle$, the spin textures are opposite for different layers for the bands $E_-$, too. Furthermore, we can find that $\langle \mathbf{S}_+^+\rangle=-\langle \mathbf{S}_+^-\rangle$ and $\langle \mathbf{S}_-^+\rangle=-\langle \mathbf{S}_-^-\rangle$, so the spin texture of the same layer are opposite for different double degenerate bands.

 The spin texture of the low energy bands in the momentum space around X point calculated by the TB model are shown in Supplementary Figure~\ref{FigS5}. We add a small asymmetric potential $U=0.01$ mV to split the double degeneracy of the bands, then the second lowest conduction band and the highest valence band are dominated by the upper layer, the lowest conduction band and the second highest valence band are dominated by the lower layer. Comparing Supplementary Figure~\ref{FigS5}(a) and (c), we can see that on the $\Gamma$--X line near the X point the spin polarization of the these two band are opposite to each other, however, on the X--M line the spin polarization of the two bands are parallel. Thus the crossing between these two bands, when we increase the asymmetric potential, on the $\Gamma$--X are protected by the mirror symmetry, and that on the X--M line are not protected. We can also see this from the coefficients of the effective model. $f_2$ and $f_3$ in the effective model determine the spin texture. $f_{2\mathrm{c}}$ and $f_{2\mathrm{v}}$ have different sign, while the $f_{3\mathrm{c}}$ and $f_{3\mathrm{v}}$ have the same sign. In addition, since $f_{2\mathrm{v}}$ and $f_{3\mathrm{v}}$ have the same sign, the spin texture of the valence bands are similar to the conventional Rashba effect, in which $f_2=f_3$. However, $f_{2\mathrm{c}}$ and $f_{3\mathrm{c}}$ have opposite sign, so the spin texture of the conduction bands is more complicated.
\begin{figure} [h]
\includegraphics[width=0.6\textwidth]{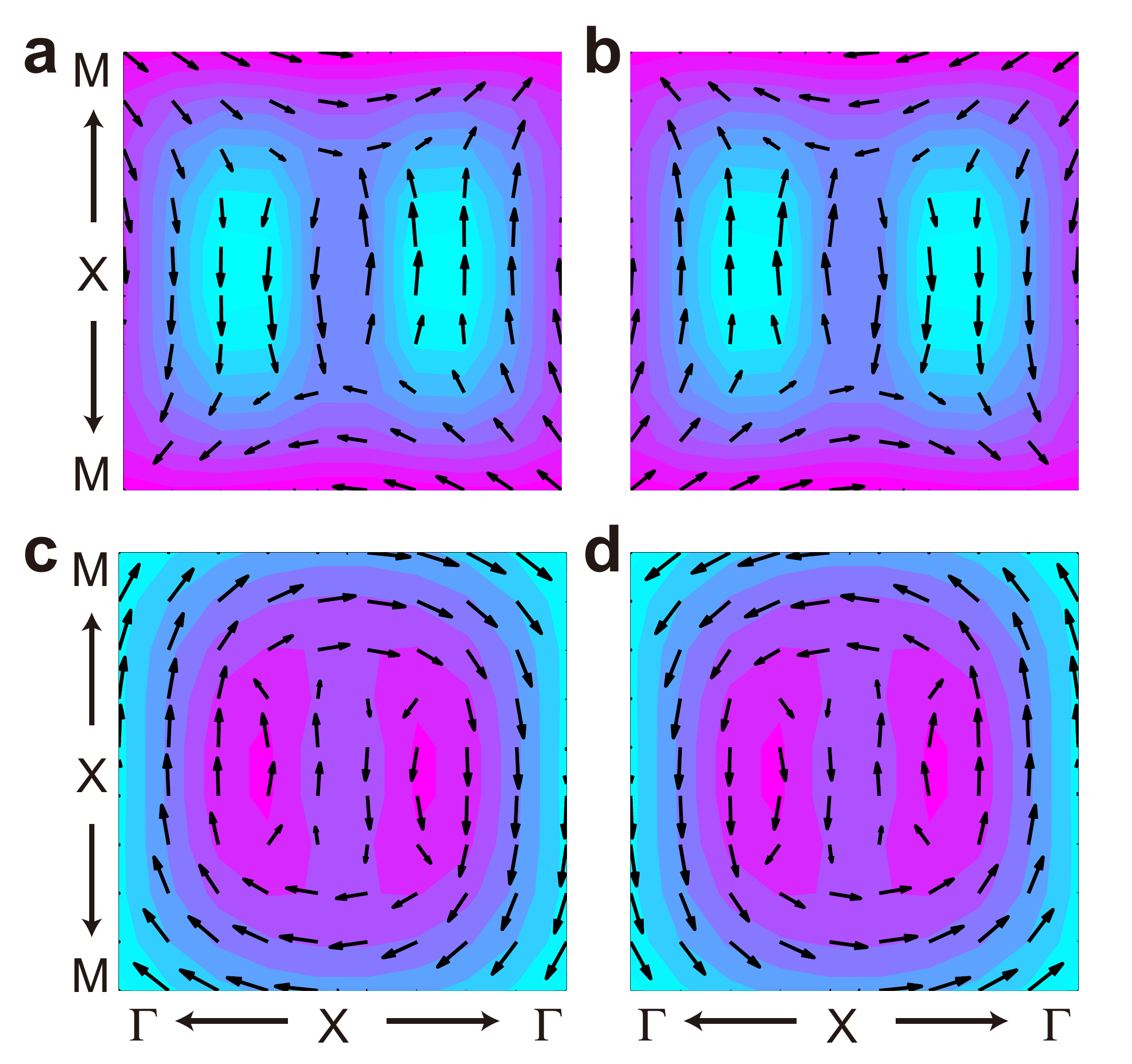}
\caption{{\bf Spin texture.} Spin texture of the low energy bands near the band gap in the momentum space around X point with a small asymmetric potential $U=0.01$ mV calculated by the TB model. (a), (b), (c) and (d) corresponds to $\mathbf{S}_-$ of the lowest conduction band, $\mathbf{S}_+$ of the second lowest conduction band, $\mathbf{S}_+$ of the highest valence band and $\mathbf{S}_-$ of the second highest valence band, respectively. The arrow represents the spin polarization. The color indicates the constant energy contours. The pink area has higher energy than the blue area in each figure. } \label{FigS5}
\end{figure}
\section{Effect of exchange coupling}
In this section, we will provide more information about the effect of the exchange coupling between electrons and magnetic moments. The energy dispersion without any asymmetric ponetial is shown in Supplementary Figure~\ref{FigS6}(a), which is the same as that in Figure 2b in the main text from the direct {\it ab initio} calculations. With an asymmetric potential, the bandgap closes and Dirac cones emerges along the $\Gamma$--X line in Supplementary Figure~\ref{FigS6}(b). When the exchange coupling is introduced, the band gap opens for Dirac cones, as shown in Supplementary Figure~\ref{FigS6}(c) and (d). This calculation confirms that the exchange coupling can induce a band gap in this system. Additional calculations of edge states have shown that the system becomes a quantum anomalous Hall insulator after opening a gap (see Figure 4 in the main text). The calculation in the main text is for (SrF)$_2$(SbSe$_2$)$_2$ and similar results are obtained for (LaO)$_2$(SbSe$_2$)$_2$ TL films, as shown in Supplementary Figure \ref{FigS7}(a)--(d). However, the electric tunability of (LaO)$_2$(SbSe$_2$)$_2$ TL films is found to be smaller than that of (SrF)$_2$(SbSe$_2$)$_2$.\\
\begin{figure} [h]
\includegraphics[width=0.6\textwidth]{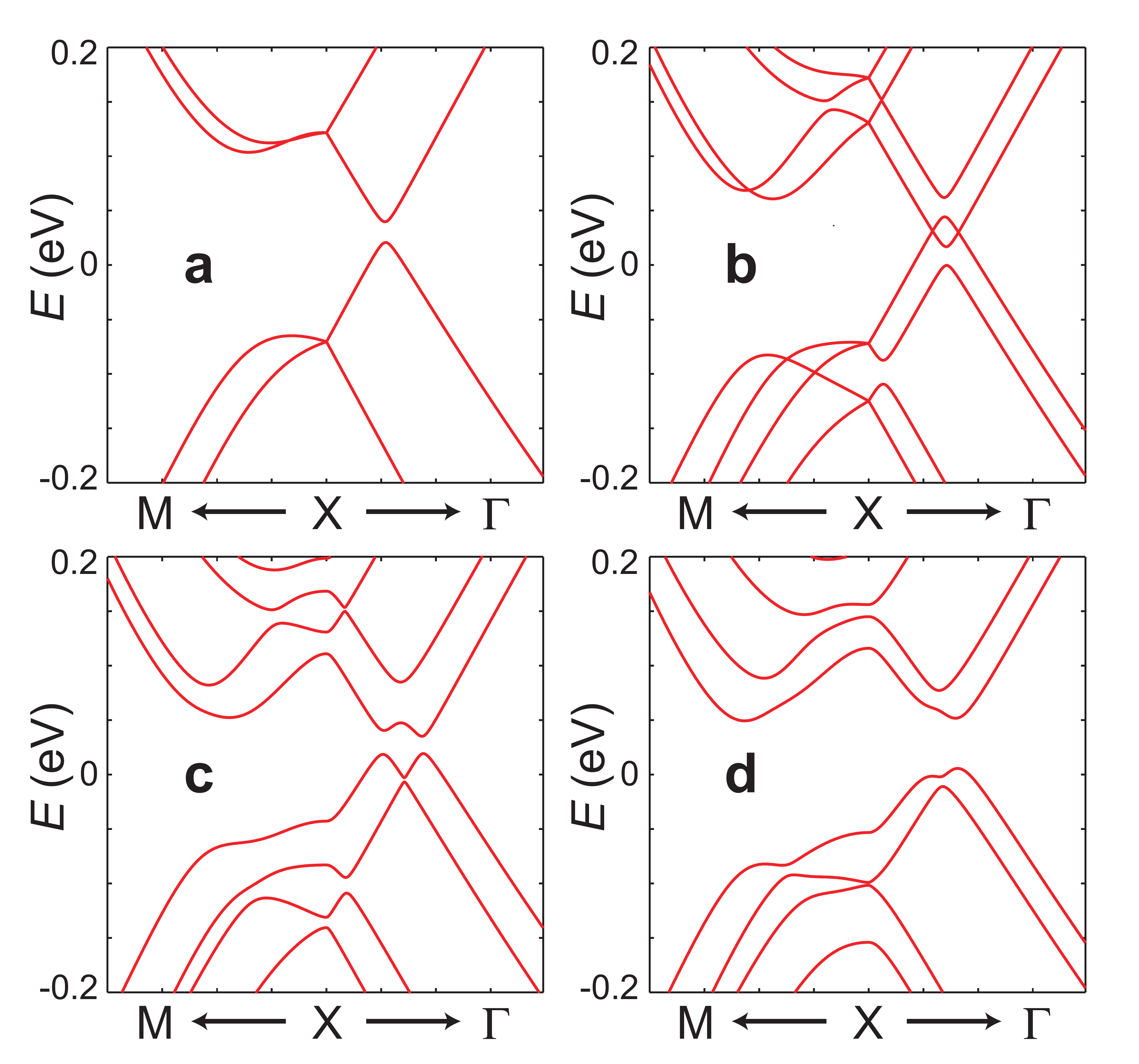}
\caption{{\bf Bulk energy dispersions.} Energy dispersion for the case (a) without asymmetric potentials and exchange coupling; (b) with the asymmetric potential e$U=0.1$ eV but no exchange coupling; (c) with the asymmetric potential e$U=0.1$ eV and the exchange coupling $J_{\mathrm{ex},\mathrm{Se}}M_{\mathrm{Se}}=0.05$ eV and $J_{\mathrm{ex},\mathrm{Sb}}M_{\mathrm{Sb}}=0$; (d) with e$U=0.1$ eV and $J_{\mathrm{ex},\mathrm{Se}}M_{\mathrm{Se}}=-J_{\mathrm{ex},\mathrm{Sb}}M_{\mathrm{Sb}}=0.05$ eV. } \label{FigS6}
\end{figure}
\begin{figure} [h]
\includegraphics[width=0.6\textwidth]{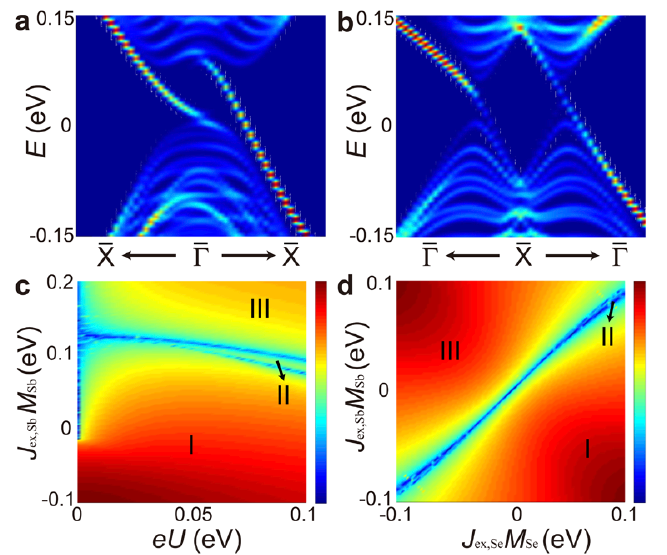}
\caption{{\bf Electrically tunable quantum anomalous Hall effect in (LaO)$_2$(SbSe$_2$)$_2$ film.} We shows the local density of states at one edge of a ribbon of (LaO)$_2$(SbSe$_2$)$_2$ film near the $\bar{\Gamma}$ point in (a) and near the $\rm{\bar{X}}$ point in (b). The bulk bandgap is plotted as a function of $J_{\mathrm{ex},\mathrm{Sb}}M_{\mathrm{Sb}}$ and e$U$ in (c) for $J_{\mathrm{ex},\mathrm{Se}}M_{\mathrm{Se}}=0.1$ eV, while it is shown as a function of $J_{\mathrm{ex},\mathrm{Sb}}M_{\mathrm{Sb}}$ and $J_{\mathrm{ex},\mathrm{Se}}M_{\mathrm{Se}}$ in (d) for e$U=0.05$ eV. The gapless lines with blue color divide the phase diagram into three regimes I, II and III with Hall conductance $\frac{4e^2}{h}$, $0$ and $-\frac{4e^2}{h}$, respectively.}  \label{FigS7}
\end{figure}
\section{The exfoliation energy and stability for a triple-layer (LaO)$_2$(SbSe$_2$)$_2$}
In this section, we discuss the experimental feasibility of the triple-layer (TL) (LaO)$_2$(SbSe$_2$)$_2$, such as the experimental preparation and the stability. Since the bulk (LaO)$_2$(SbSe$_2$)$_2$ has been prepared in experiment\cite{guittard1984}, we expect that a single TL of (LaO)$_2$(SbSe$_2$)$_2$ can be exfoliated from its bulk. Here we calculate the exfoliation energy between two TLs. The energy vs. interlayer spacing is shown in Supplementary Figure~\ref{FigS9}. It can be seen that the binding energy between two TLs for (LaO)$_2$(SbSe$_2$)$_2$ is about 7.3 meV {\AA}$^{-2}$ without van der Waals (vdW) correction and 30.7 meV {\AA}$^{-2}$ with vdW correction. As a comparison, we notice that the interlayer binding energy for MoS$_2$\cite{bjorkman2012} is around 20 meV {\AA}$^{-2}$ and the corresponding single layer film has been produced by exfoliation in experiments\cite{novoselov2005two,coleman2011two}. The binding energy for monolayer (LaO)$_2$(SbSe$_2$)$_2$ is larger compared to that of MoS$_2$, but we expect the exfoliation of monolayer (LaO)$_2$(SbSe$_2$)$_2$ is still feasible in experiments. The binding energy between the triple layers for another material (SrF)$_2$(SbSe$_2$)$_2$ has the similar values: 8.7 meV {\AA}$^{-2}$ without vdW correction and 31.8 meV {\AA}$^{-2}$ with vdW correction.

About the stability of 1-TL (LaO)$_2$(SbSe$_2$)$_2$, we calculate the phonon spectrum by the direct finite displacement method with VASP, and show it in Supplementary Figure~\ref{FigS10}(a). We find negative frequencies around $\Gamma$ and M points. The crystal instability in this class of materials has been studied for (LaO)$_2$(BiS$_2$)$_2$\cite{yildirim2013}, but this issue is quite controversial. For example, some calculations have shown that all the branches of the phonon spectrum have positive frequencies with no negative phonon modes by considering $2 \times 2$ supercell, suggesting the dynamic stability of the 1-TL (LaO)$_2$(BiS$_2$)$_2$\cite{liu2013}. While some other calculations show that instability indeed exists in (LaO)$_2$(BiS$_2$)$_2$, leading to lattice distortion that lowers crystal symmetry\cite{yildirim2013}. Further argument suggests that finite temperature-dependent entropy may drive the stabilization of this class of materials\cite{souvatzis2008,yildirim2013}. In addition, the monolayer material is usually sustained on the substrate, and may be stabilized by choosing appropriate substrates and delicately controlling the growth kinetics. Therefore, a simple phonon spectrum calculation is difficult to convincingly determine the stability of the lattice and the type of distortion if it exists.

We would like to emphasize that even if taking into account the distortion, Dirac physics can still exist and depends on the type of distortion. For example, for (LaO)$_2$(BiS$_2$)$_2$ without any doping, it is shown that distortion can lower the symmetry group from $P4/nmm$ to $P21mn$\cite{yildirim2013}, as shown in Supplementary Figure~\ref{FigS10}(b). Since our gapless Dirac cones are protected by mirror symmetry, we expect a mass is introduced when mirror symmetry is broken, resulting in a massive Dirac equation for each Dirac cone. For the distortion in Supplementary Figure~\ref{FigS10}(b), the mirror symmetry in one direction $m_y$ is broken, while another mirror symmetry $m_x$ is preserved. The gapless Dirac cones around the X point is protected by $m_y$, and thus with this distortion they should be gapped. However, since $m_x$ is preserved, the Dirac cones around the Y point should be still gapless. To demonstrate this, we perform the first-principles calculations with a distortion of 0.12 {\AA} along $y$ for the Se$_2$/Se$^{\prime}_2$ atoms. Results shown in Supplementary Figure~\ref{FigS10}(c) and (d) confirm the analysis above. Note that the gap along the X--$\Gamma$ direction is so tiny (about 4 meV) that it is difficult to distinguish it from numerical errors, which indicates that the Dirac cones are insensitive to the lattice distortion. In Supplementary Figure~\ref{FigS10}(e) and (f), TB calculations clearly reveal the gapped Dirac cones around X point and gapless Dirac cones around Y point. However, it is emphasized that the electrically tunable quantum anomalous Hall effect will not be affected by this distortion once the magnetization gap is larger than the distortion-induced gap.

We have also performed the lattice relaxation of a TL structure. Compared with the experimental 4.13 {\AA}, the optimal lattice constant is 4.09 {\AA}. There are some quantitative changes on the band dispersion for the relaxed structure. As shown in Supplementary Figure~\ref{FigS13}, the gap changes from 67 meV to 9 meV without the external electric field, and thus a smaller electric field is enough to close the band gap and drives the system into the gapless Dirac cone phase. Under the same electric field (as long as not too large), the relaxed structure has a larger Dirac Fermi velocity along the X--$\Gamma$ direction than the value of experimental structure. Despite these quantitative changes, we still emphasize that the main result of electrically tunable multiple Dirac cones is not influenced by the structure relaxation.
\begin{figure} [h]
\includegraphics[width=0.6\textwidth]{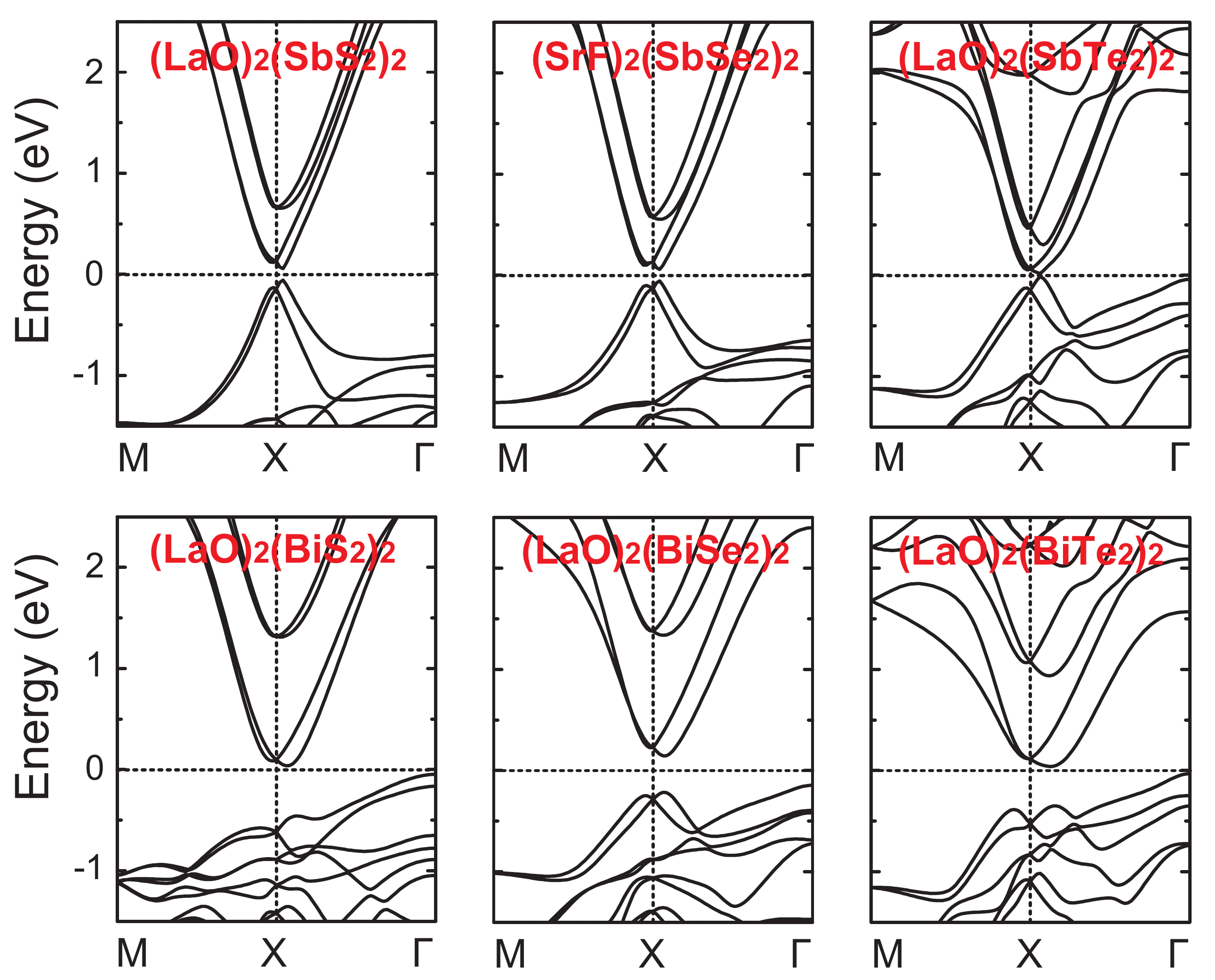}
\caption{{\bf Bulk band structures of other (LaO)$_2$(SbSe$_2$)$_2$ class of materials.}} \label{FigS2}
\end{figure}
\begin{figure} [h]
\includegraphics[width=0.6\textwidth]{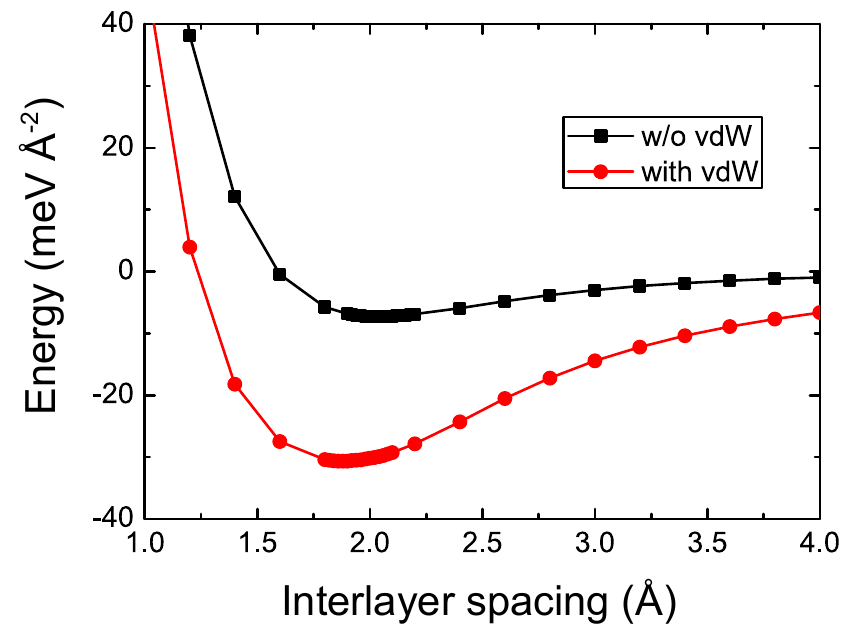}
\caption{{\bf Binding-energy curve for two TLs.} The black line is the calculation without van der Waals (vdW) correction, and the red line includes vdW correction.}  \label{FigS9}
\end{figure}
\begin{figure} [h]
\includegraphics[width=0.6\textwidth]{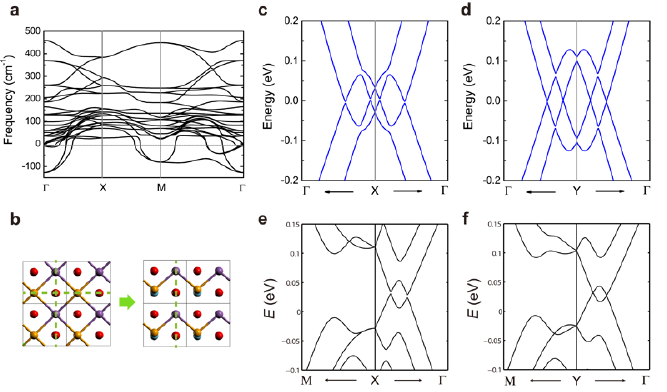}
\caption{{\bf Phonon spectrum and distortion effects.} (a)The phonon spectrum of 1-TL (LaO)$_2$(SbSe$_2$)$_2$. (b)Schematic plot of the lattice distortion in the top view. The symmetry groups of the crystal structure without distortion (left) and with distortion (right) are, respectively, $P4/nmm$ and $P21mn$. The green dashed lines indicate the mirror symmetry. (c) and (d), the band structures around the X and Y points, respectively, calculated by the first-principles calculations with a distortion of 0.12 {\AA} along $y$ for the Se$_2$/Se$^{\prime}_2$ atoms. The Dirac cones gain a tiny gap along the X--$\Gamma$ line, but remain gapless along the Y--$\Gamma$ line. (e) and (f), the band structures around the X and Y points calculated by the TB model with the distortion shown in (b).}  \label{FigS10}
\end{figure}
\begin{figure} [h]
\includegraphics[width=0.6\textwidth]{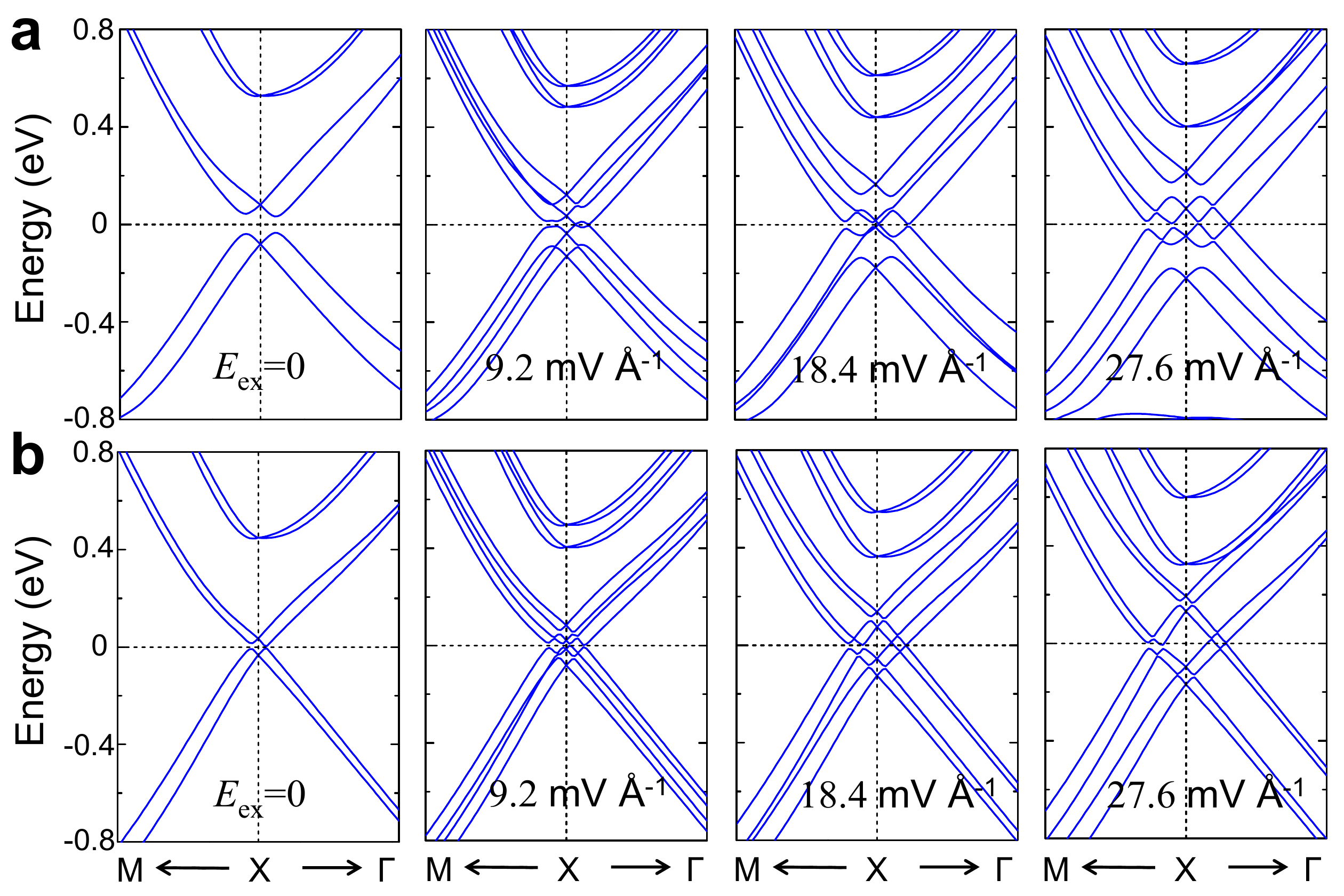}
\caption{{\bf The effect of lattice relaxation.} Band dispersions of a TL film near X under different electric fields with the experimental lattice (a) and the relaxed lattice (b).}  \label{FigS13}
\end{figure}
\section{Dirac physics in multiple-TLs (LaO)$_2$(SbSe$_2$)$_2$}
The interlayer coupling in multiple-layer systems is inevitable and leads to strong hybridization between low energy states in different layers. However, it will not change the electrical tunability of Dirac physics discussed here. To clearly illustrate this situation, we take the two triple layers as an example [Supplementary Figure~\ref{FigS11}(a)]. The projected bands are shown in Supplementary Figure~\ref{FigS11}(b)-(d). The low energy states are still from the contribution of four SbSe layers. Without external electric field [Supplementary Figure~\ref{FigS11}(b)], the system has both the inversion symmetry and time reversal symmetry. Therefore, each band is double degenerate and consists of energy states from different SbSe layers. For example, the red curves of dispersion in Supplementary Figure~\ref{FigS11}(b) consist of the energy states from the top (red dot in Supplementary Figure~\ref{FigS11}(a)) and bottom (pink dot) SbSe layers, while the blue curves in Supplementary Figure~\ref{FigS11}(b) consist of the states from the second (blue dot) and third (green dot) SbSe layers. The strong interlayer coupling results in large splitting between the red and blue curves of energy dispersion. The gaps of the blue bands and red bands are also different, estimated as 13 meV and 69 meV, respectively. When we apply an external electric field of 27.6 mV/{\AA} on this two TLs [Supplementary Figure~\ref{FigS11}(c) and (d)], each double degenerate band splits into two bands, which possess the opposite spin texture, similar to the single TL case. We still find gapless crossing points along the $\Gamma$--X line [the labelled A and B points in Supplementary Figure~\ref{FigS11}(c) and (d)], while those along the M--X line are gapped [the labelled C and D points in Supplementary Figure~\ref{FigS11}(c) and (d)]. The main difference between the single TL and multiple TLs lies in the fact that more bands exist in multiple TLs so the gapless Dirac cones may be buried in other metallic modes. Nevertheless, we emphasize that the similar Dirac physics with the electrical tunability can exist in the multiple-layer systems.

\begin{figure} [h]
\includegraphics[width=0.6\textwidth]{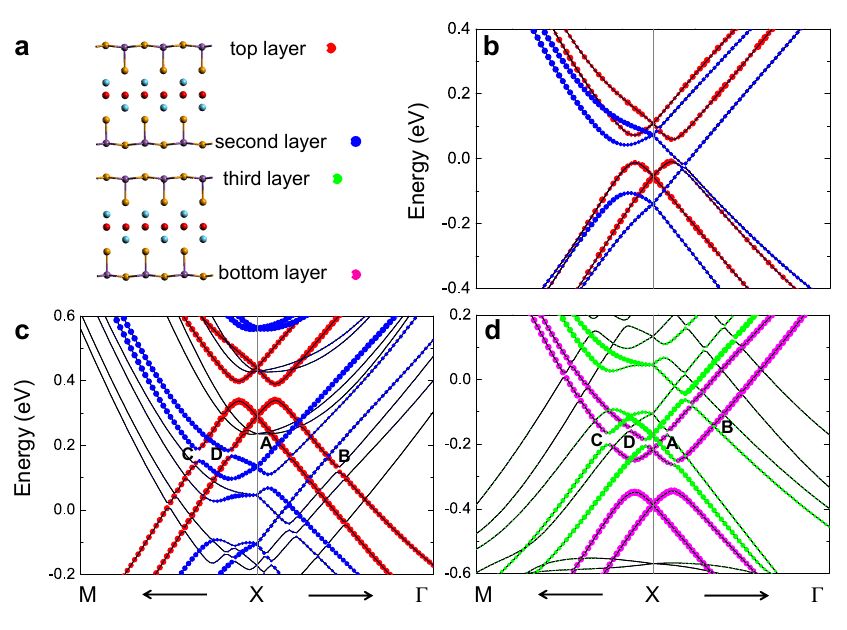}
\caption{{\bf Band structures of two triple layers of (LaO)$_2$(SbSe$_2$)$_2$.} (a) without the external electric field (b) and with the external electric field of 27.6 mV {\AA}$^{-1}$ (c, d). The red (blue/green/pink) circle in (b)--(d) represents the weight of the top (second/third/bottom) SbSe layer. Note that we adopt different energy scales for (b)--(d).}  \label{FigS11}
\end{figure}

\section{The strain effect}
To prove that the Dirac cones are indeed protected by the mirror symmetry, we add a strain as a perturbation into the TB model around X(Y) point to break the mirror symmetry $m_x$ and $m_y$, and study the effects of strain on the band structure around X(Y) point. An uniaxial stress in (110) direction can be added by a small nonzero term $\epsilon_{xy}$ in the strain tensor. The results are shown in Supplementary Figure~\ref{FigS8}. The gapless Dirac cones on X(Y)-$\Gamma$ around X(Y) point are gapped in Supplementary Figure~\ref{FigS8}(a) and (b) with $\epsilon_{xy}=0.1$ and asymmetric potential $U=80$ mV. Therefore, we have proved that the Dirac cones are protected by the mirror symmetry.
\begin{figure} [h]
\includegraphics[width=0.6\textwidth]{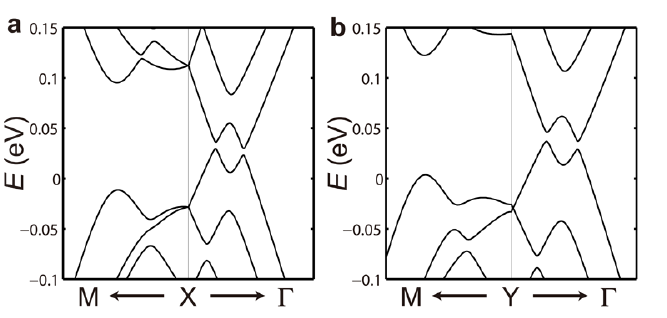}
\caption{{\bf Strain effects on band structure.} Band structure around X  and Y point on the high symmetry lines with uniaxial strain in (110) direction in (a) and (b), respectively, where $\epsilon_{xy}=0.1$ and asymmetric potential $U=80$mV.} \label{FigS8}
\end{figure}
\\
\\
\\
\\
{\bf Supplementary References}

\end{widetext}


\begin{thebibliography}{10}
\expandafter\ifx\csname url\endcsname\relax
  \def\url#1{\texttt{#1}}\fi
\expandafter\ifx\csname urlprefix\endcsname\relax\def\urlprefix{URL }\fi
\providecommand{\bibinfo}[2]{#2}
\providecommand{\eprint}[2][]{\url{#2}}

\bibitem{neto2009}
\bibinfo{author}{Neto, A.~C.}, \bibinfo{author}{Guinea, F.},
  \bibinfo{author}{Peres, N.}, \bibinfo{author}{Novoselov, K.~S.} \&
  \bibinfo{author}{Geim, A.~K.}
\newblock \bibinfo{title}{The electronic properties of graphene}.
\newblock \emph{\bibinfo{journal}{Reviews of modern physics}}
  \textbf{\bibinfo{volume}{81}}, \bibinfo{pages}{109--162} (\bibinfo{year}{2009}).

\bibitem{hasan2010}
\bibinfo{author}{Hasan, M.~Z.} \& \bibinfo{author}{Kane, C.~L.}
\newblock \bibinfo{title}{Colloquium: topological insulators}.
\newblock \emph{\bibinfo{journal}{Reviews of Modern Physics}}
  \textbf{\bibinfo{volume}{82}}, \bibinfo{pages}{3045--3067} (\bibinfo{year}{2010}).

\bibitem{qi2011}
\bibinfo{author}{Qi, X.-L.} \& \bibinfo{author}{Zhang, S.-C.}
\newblock \bibinfo{title}{Topological insulators and superconductors}.
\newblock \emph{\bibinfo{journal}{Reviews of Modern Physics}}
  \textbf{\bibinfo{volume}{83}}, \bibinfo{pages}{1057--1110} (\bibinfo{year}{2011}).

\bibitem{hsieh2012}
\bibinfo{author}{Hsieh, T.~H.}, \bibinfo{author}{Lin, H.},
  \bibinfo{author}{Liu, J.}, \bibinfo{author}{Duan, W.}, \bibinfo{author}{Bansil, A.} \& \bibinfo{author}{Fu, L.}
\newblock \bibinfo{title}{Topological crystalline insulators in the SnTe
  material class}.
\newblock \emph{\bibinfo{journal}{Nature communications}}
  \textbf{\bibinfo{volume}{3}}, \bibinfo{pages}{982} (\bibinfo{year}{2012}).

\bibitem{tanaka2012}
\bibinfo{author}{Tanaka, Y.} \emph{et~al.}
\newblock \bibinfo{title}{Experimental realization of a topological crystalline
  insulator in SnTe}.
\newblock \emph{\bibinfo{journal}{Nature Physics}}
  \textbf{\bibinfo{volume}{8}}, \bibinfo{pages}{800--803}
  (\bibinfo{year}{2012}).

\bibitem{dziawa2012}
\bibinfo{author}{Dziawa, P.} \emph{et~al.}
\newblock \bibinfo{title}{Topological crystalline insulator states in Pb$_{1-
x}$Sn$_x$Se}.
\newblock \emph{\bibinfo{journal}{Nature materials}}
  \textbf{\bibinfo{volume}{11}}, \bibinfo{pages}{1023--1027}
  (\bibinfo{year}{2012}).

\bibitem{splendiani2010}
\bibinfo{author}{Splendiani, A.} \emph{et~al.}
\newblock \bibinfo{title}{Emerging photoluminescence in monolayer MoS$_2$}.
\newblock \emph{\bibinfo{journal}{Nano letters}} \textbf{\bibinfo{volume}{10}},
  \bibinfo{pages}{1271--1275} (\bibinfo{year}{2010}).

\bibitem{mak2010}
\bibinfo{author}{Mak, K.~F.}, \bibinfo{author}{Lee, C.}, \bibinfo{author}{Hone,
  J.}, \bibinfo{author}{Shan, J.} \& \bibinfo{author}{Heinz, T.~F.}
\newblock \bibinfo{title}{Atomically thin MoS$_2$: a new direct-gap
  semiconductor}.
\newblock \emph{\bibinfo{journal}{Physical Review Letters}}
  \textbf{\bibinfo{volume}{105}}, \bibinfo{pages}{136805}
  (\bibinfo{year}{2010}).

\bibitem{xiao2012}
\bibinfo{author}{Xiao, D.}, \bibinfo{author}{Liu, G.-B.},
  \bibinfo{author}{Feng, W.}, \bibinfo{author}{Xu, X.} \& \bibinfo{author}{Yao,
  W.}
\newblock \bibinfo{title}{Coupled spin and valley physics in monolayers of MoS$_2$ and other group-VI dichalcogenides}.
\newblock \emph{\bibinfo{journal}{Physical Review Letters}}
  \textbf{\bibinfo{volume}{108}}, \bibinfo{pages}{196802}
  (\bibinfo{year}{2012}).

\bibitem{cao2012}
\bibinfo{author}{Cao, T.} \emph{et~al.}
\newblock \bibinfo{title}{Valley-selective circular dichroism of monolayer
  Molybdenum Disulphide}.
\newblock \emph{\bibinfo{journal}{Nature communications}}
  \textbf{\bibinfo{volume}{3}}, \bibinfo{pages}{887} (\bibinfo{year}{2012}).

\bibitem{zeng2012}
\bibinfo{author}{Zeng, H.}, \bibinfo{author}{Dai, J.}, \bibinfo{author}{Yao,
  W.}, \bibinfo{author}{Xiao, D.} \& \bibinfo{author}{Cui, X.}
\newblock \bibinfo{title}{Valley polarization in MoS$_2$ monolayers by optical
  pumping}.
\newblock \emph{\bibinfo{journal}{Nature nanotechnology}}
  \textbf{\bibinfo{volume}{7}}, \bibinfo{pages}{490--493}
  (\bibinfo{year}{2012}).

\bibitem{park2011}
\bibinfo{author}{Park, J.} \emph{et~al.}
\newblock \bibinfo{title}{Anisotropic Dirac fermions in a Bi square net of
  SrMnBi$_2$}.
\newblock \emph{\bibinfo{journal}{Physical review letters}}
  \textbf{\bibinfo{volume}{107}}, \bibinfo{pages}{126402}
  (\bibinfo{year}{2011}).

\bibitem{tsuei2000}
\bibinfo{author}{Tsuei, C.} \& \bibinfo{author}{Kirtley, J.}
\newblock \bibinfo{title}{Pairing symmetry in cuprate superconductors}.
\newblock \emph{\bibinfo{journal}{Reviews of Modern Physics}}
  \textbf{\bibinfo{volume}{72}}, \bibinfo{pages}{969} (\bibinfo{year}{2000}).

\bibitem{liu2014}
\bibinfo{author}{Liu, Z.} \emph{et~al.}
\newblock \bibinfo{title}{Discovery of a three-dimensional topological dirac
  semimetal, Na$_3$Bi}.
\newblock \emph{\bibinfo{journal}{Science}} \textbf{\bibinfo{volume}{343}},
  \bibinfo{pages}{864--867} (\bibinfo{year}{2014}).

\bibitem{borisenko2013}
\bibinfo{author}{Borisenko, S.} \emph{et~al.}
\newblock \bibinfo{title}{Experimental realization of a three-dimensional Dirac
  semimetal}.
\newblock \emph{\bibinfo{journal}{Physical review letters}}
  \textbf{\bibinfo{volume}{113}}, \bibinfo{pages}{027603}
  (\bibinfo{year}{2014}).

\bibitem{neupane2014}
\bibinfo{author}{Neupane, M.} \emph{et~al.}
\newblock \bibinfo{title}{Observation of a three-dimensional topological Dirac
  semimetal phase in high-mobility Cd$_3$As$_2$}.
\newblock \emph{\bibinfo{journal}{Nature communications}}
  \textbf{\bibinfo{volume}{5}}, \bibinfo{pages}{3786} (\bibinfo{year}{2014}).

\bibitem{wehling2014}
\bibinfo{author}{Wehling, T.}, \bibinfo{author}{Black-Schaffer, A.} \&
  \bibinfo{author}{Balatsky, A.}
\newblock \bibinfo{title}{Dirac materials}.
\newblock \emph{\bibinfo{journal}{Advances in Physics}}
  \textbf{\bibinfo{volume}{63}}, \bibinfo{pages}{1--76}
  (\bibinfo{year}{2014}).

\bibitem{zhang2009a}
\bibinfo{author}{Zhang, Y.} \emph{et~al.}
\newblock \bibinfo{title}{Direct observation of a widely tunable bandgap in
  bilayer graphene}.
\newblock \emph{\bibinfo{journal}{Nature}} \textbf{\bibinfo{volume}{459}},
  \bibinfo{pages}{820--823} (\bibinfo{year}{2009}).

\bibitem{xia2010}
\bibinfo{author}{Xia, F.}, \bibinfo{author}{Farmer, D.~B.},
  \bibinfo{author}{Lin, Y.-m.} \& \bibinfo{author}{Avouris, P.}
\newblock \bibinfo{title}{Graphene field-effect transistors with high on/off
  current ratio and large transport band gap at room temperature}.
\newblock \emph{\bibinfo{journal}{Nano letters}} \textbf{\bibinfo{volume}{10}},
  \bibinfo{pages}{715--718} (\bibinfo{year}{2010}).

\bibitem{guittard1984}
\bibinfo{author}{Guittard, M.} \emph{et~al.}
\newblock \bibinfo{title}{Oxysulfides and oxyselenides in sheets, formed by a
  rare earth element and a second metal}.
\newblock \emph{\bibinfo{journal}{Journal of Solid State Chemistry}}
  \textbf{\bibinfo{volume}{51}}, \bibinfo{pages}{227--238}
  (\bibinfo{year}{1984}).

\bibitem{yazici2013}
\bibinfo{author}{Yazici, D.}, \bibinfo{author}{Huang, K.},
  \bibinfo{author}{White, B.}, \bibinfo{author}{Chang, A.}, \bibinfo{author}{Friedman, A.} \& \bibinfo{author}{Maple, M.}
\newblock \bibinfo{title}{Superconductivity of F-substituted LnOBiS$_2$ (Ln= La,
  Ce, Pr, Nd, Yb) compounds}.
\newblock \emph{\bibinfo{journal}{Philosophical Magazine}}
  \textbf{\bibinfo{volume}{93}}, \bibinfo{pages}{673--680}
  (\bibinfo{year}{2013}).

\bibitem{mizuguchi2012}
\bibinfo{author}{Mizuguchi, Y.} \emph{et~al.}
\newblock \bibinfo{title}{Superconductivity in novel BiS$_2$-based layered
superconductor LaO$_{1-x}$F$_x$BiS$_2$}.
\newblock \emph{\bibinfo{journal}{Journal of the Physical Society of Japan}}
  \textbf{\bibinfo{volume}{81}}, \bibinfo{pages}{114725} (\bibinfo{year}{2012}).

\bibitem{demura2013}
\bibinfo{author}{Demura, S.} \emph{et~al.}
\newblock \bibinfo{title}{New member of BiS$_2$-based superconductor NdO$_{1-x}$F$_x$BiS$_2$}.
\newblock \emph{\bibinfo{journal}{Journal of the Physical Society of Japan}}
  \textbf{\bibinfo{volume}{82}}, \bibinfo{pages}{033708} (\bibinfo{year}{2013}).

\bibitem{kabbour2006}
\bibinfo{author}{Kabbour, H.} \& \bibinfo{author}{Cario, L.}
\newblock \bibinfo{title}{Ae$_2$Sb$_2$X$_4$F$_2$ (Ae= Sr, Ba): New members of the
homologous series Ae$_2$M$_{1+n}$ X$_{3+n}$ F$_2$ designed from rock salt and fluorite 2D
  building blocks}.
\newblock \emph{\bibinfo{journal}{Inorganic chemistry}}
  \textbf{\bibinfo{volume}{45}}, \bibinfo{pages}{2713--2717}
  (\bibinfo{year}{2006}).

\bibitem{lei2013}
\bibinfo{author}{Lei, H.}, \bibinfo{author}{Wang, K.},
  \bibinfo{author}{Abeykoon, M.}, \bibinfo{author}{Bozin, E.~S.} \&
  \bibinfo{author}{Petrovic, C.}
\newblock \bibinfo{title}{New layered fluorosulfide SrFBiS$_2$}.
\newblock \emph{\bibinfo{journal}{Inorganic chemistry}}
  \textbf{\bibinfo{volume}{52}}, \bibinfo{pages}{10685--10689}
  (\bibinfo{year}{2013}).

\bibitem{lin2013}
\bibinfo{author}{Lin, X.} \emph{et~al.}
\newblock \bibinfo{title}{Superconductivity induced by La doping in Sr$_{1- x}$ La$_x$ FBiS$_2$}.
\newblock \emph{\bibinfo{journal}{Physical Review B}}
  \textbf{\bibinfo{volume}{87}}, \bibinfo{pages}{020504}
  (\bibinfo{year}{2013}).

\bibitem{johnson1974}
\bibinfo{author}{Johnson, V.} \& \bibinfo{author}{Jeitschko, W.}
\newblock \bibinfo{title}{ZrCuSiAs: A ¡°filled¡± PbFCl type}.
\newblock \emph{\bibinfo{journal}{Journal of Solid State Chemistry}}
  \textbf{\bibinfo{volume}{11}}, \bibinfo{pages}{161--166}
  (\bibinfo{year}{1974}).

\bibitem{stewart2011}
\bibinfo{author}{Stewart, G.}
\newblock \bibinfo{title}{Superconductivity in iron compounds}.
\newblock \emph{\bibinfo{journal}{Reviews of Modern Physics}}
  \textbf{\bibinfo{volume}{83}}, \bibinfo{pages}{1589--1652} (\bibinfo{year}{2011}).

\bibitem{liu2013}
\bibinfo{author}{Liu, Q.}, \bibinfo{author}{Guo, Y.} \&
  \bibinfo{author}{Freeman, A.~J.}
\newblock \bibinfo{title}{Tunable rashba effect in two-dimensional LaOBiS$_2$
  films: Ultrathin candidates for spin field effect transistors}.
\newblock \emph{\bibinfo{journal}{Nano letters}} \textbf{\bibinfo{volume}{13}},
  \bibinfo{pages}{5264--5270} (\bibinfo{year}{2013}).

\bibitem{novoselov2004}
\bibinfo{author}{Novoselov, K.~S.} \emph{et~al.}
\newblock \bibinfo{title}{Electric field effect in atomically thin carbon
  films}.
\newblock \emph{\bibinfo{journal}{science}} \textbf{\bibinfo{volume}{306}},
  \bibinfo{pages}{666--669} (\bibinfo{year}{2004}).

\bibitem{castro2007}
\bibinfo{author}{Castro, E.~V.} \emph{et~al.}
\newblock \bibinfo{title}{Biased bilayer graphene: semiconductor with a gap
  tunable by the electric field effect}.
\newblock \emph{\bibinfo{journal}{Physical Review Letters}}
  \textbf{\bibinfo{volume}{99}}, \bibinfo{pages}{216802}
  (\bibinfo{year}{2007}).

\bibitem{winkler2003}
\bibinfo{author}{Winkler, R.}
\newblock \emph{\bibinfo{title}{Spin-Orbit Coupling Effects in Two-Dimensional
  Electron and Hole Systems}} (\bibinfo{publisher}{Springer},
  \bibinfo{year}{2003}).

\bibitem{zhang2014a}
\bibinfo{author}{Zhang, X.}, \bibinfo{author}{Liu, Q.}, \bibinfo{author}{Luo,
  J.-W.}, \bibinfo{author}{Freeman, A.~J.} \& \bibinfo{author}{Zunger, A.}
\newblock \bibinfo{title}{Hidden spin polarization in inversion-symmetric bulk
  crystals}.
\newblock \emph{\bibinfo{journal}{Nature Physics}}
  \textbf{\bibinfo{volume}{10}}, \bibinfo{pages}{387--393}
  (\bibinfo{year}{2014}).

\bibitem{haldane1988}
\bibinfo{author}{Haldane, F. D. M.}
  \newblock \bibinfo{title}{Model for a quantum Hall effect without Landau levels: Condensed-matter realization of the ``parity anomaly"}.
\newblock \emph{\bibinfo{journal}{Physical review letters}}
  \textbf{\bibinfo{volume}{61}}, \bibinfo{pages}{2015}
  (\bibinfo{year}{1988}).

\bibitem{onoda2003}
\bibinfo{author}{Onoda, M.} \& \bibinfo{author}{Nagaosa, N.}
  \newblock \bibinfo{title}{Quantized anomalous Hall effect in two-dimensional ferromagnets: quantum Hall effect in metals}.
\newblock \emph{\bibinfo{journal}{Physical review letters}}
  \textbf{\bibinfo{volume}{90}}, \bibinfo{pages}{206601}
  (\bibinfo{year}{2003}).

\bibitem{qi2006}
\bibinfo{author}{Qi, X.-L.}, \bibinfo{author}{Wu, Y.-S.} \& \bibinfo{author}{Zhang, S.-C.}
\newblock \bibinfo{title}{Topological quantization of the spin Hall effect in two-dimensional paramagnetic semiconductors}.
\newblock \emph{\bibinfo{journal}{Physical review B}}
  \textbf{\bibinfo{volume}{74}}, \bibinfo{pages}{085308} (\bibinfo{year}{2006}).

\bibitem{liu2008}
\bibinfo{author}{Liu, C.-X.}, \bibinfo{author}{Qi, X.-L.},
  \bibinfo{author}{Dai, X.}, \bibinfo{author}{Fang, Z.} \&
  \bibinfo{author}{Zhang, S.-C.}
  \newblock \bibinfo{title}{Quantum anomalous Hall effect in Hg$_{1-y}$Mn$_y$Te  quantum wells}.
\newblock \emph{\bibinfo{journal}{Physical review letters}}
  \textbf{\bibinfo{volume}{101}}, \bibinfo{pages}{146802}
  (\bibinfo{year}{2008}).

\bibitem{yu2010}
\bibinfo{author}{Yu, R.}, \bibinfo{author}{Zhang, W.}, \bibinfo{author}{Zhang, H.-J.}, \bibinfo{author}{Zhang, S.-C.}, \bibinfo{author}{Dai, X.} \&
  \bibinfo{author}{Fang, Z.}
\newblock \bibinfo{title}{Quantized anomalous Hall effect in magnetic
  topological insulators}.
\newblock \emph{\bibinfo{journal}{Science}} \textbf{\bibinfo{volume}{329}},
  \bibinfo{pages}{61--64} (\bibinfo{year}{2010}).

\bibitem{chang2013}
\bibinfo{author}{Chang, C.-Z.} \emph{et~al.}
\newblock \bibinfo{title}{Experimental observation of the quantum anomalous
  Hall effect in a magnetic topological insulator}.
\newblock \emph{\bibinfo{journal}{Science}} \textbf{\bibinfo{volume}{340}},
  \bibinfo{pages}{167--170} (\bibinfo{year}{2013}).

\bibitem{chang2015}
\bibinfo{author}{Chang, C.-Z.} \emph{et~al.}
\newblock \bibinfo{title}{High-precision realization of robust quantum anomalous Hall state in a hard ferromagnetic topological insulator}.
\newblock \emph{\bibinfo{journal}{Nature Materials}}
  \textbf{\bibinfo{volume}{14}}, \bibinfo{pages}{473--477}
  (\bibinfo{year}{2015}).

\bibitem{qiao2010}
\bibinfo{author}{Qiao, Z.} \emph{et~al.}
\newblock \bibinfo{title}{Quantum anomalous Hall effect in graphene from Rashba and exchange effects}.
\newblock \emph{\bibinfo{journal}{Physical review B}}
  \textbf{\bibinfo{volume}{82}}, \bibinfo{pages}{161414} (\bibinfo{year}{2010}).

\bibitem{jiang2012}
\bibinfo{author}{Jiang, H.}, \bibinfo{author}{Qiao, Z.}, \bibinfo{author}{Liu, H.} \& \bibinfo{author}{Niu, Q.}
\newblock \bibinfo{title}{Quantum anomalous Hall effect with tunable Chern number in magnetic topological insulator film}.
\newblock \emph{\bibinfo{journal}{Physical review B}}
  \textbf{\bibinfo{volume}{85}}, \bibinfo{pages}{045445} (\bibinfo{year}{2012}).

\bibitem{fang2014}
\bibinfo{author}{Fang, C.}, \bibinfo{author}{Gilbert, M. J.} \& \bibinfo{author}{Bernevig, B. A.}
\newblock \bibinfo{title}{Large-Chern-Number Quantum Anomalous Hall Effect in Thin-Film Topological Crystalline Insulators}.
\newblock \emph{\bibinfo{journal}{Physical Review Letters}}
  \textbf{\bibinfo{volume}{112}}, \bibinfo{pages}{046801}
  (\bibinfo{year}{2014}).

\bibitem{marzari1997}
\bibinfo{author}{Marzari, N.} \& \bibinfo{author}{Vanderbilt, D.}
\newblock \bibinfo{title}{Maximally localized generalized wannier functions for
  composite energy bands}.
\newblock \emph{\bibinfo{journal}{Physical review B}}
  \textbf{\bibinfo{volume}{56}}, \bibinfo{pages}{12847} (\bibinfo{year}{1997}).

\bibitem{souza2001}
\bibinfo{author}{Souza, I.}, \bibinfo{author}{Marzari, N.} \&
  \bibinfo{author}{Vanderbilt, D.}
\newblock \bibinfo{title}{Maximally localized wannier functions for entangled
  energy bands}.
\newblock \emph{\bibinfo{journal}{Physical Review B}}
  \textbf{\bibinfo{volume}{65}}, \bibinfo{pages}{035109}
  (\bibinfo{year}{2001}).

\bibitem{wang2013}
\bibinfo{author}{Wang, J.}, \bibinfo{author}{Lian, B.}, \bibinfo{author}{Zhang,
  H.}, \bibinfo{author}{Xu, Y.} \& \bibinfo{author}{Zhang, S.-C.}
\newblock \bibinfo{title}{Quantum anomalous Hall effect with higher plateaus}.
\newblock \emph{\bibinfo{journal}{Phys. Rev. Lett.}}
  \textbf{\bibinfo{volume}{111}}, \bibinfo{pages}{136801}
  (\bibinfo{year}{2013}).

\bibitem{usui2012}
\bibinfo{author}{Usui, H.}, \bibinfo{author}{Suzuki, K.} \&
  \bibinfo{author}{Kuroki, K.}
\newblock \bibinfo{title}{Minimal electronic models for superconducting BiS$_2$
  layers}.
\newblock \emph{\bibinfo{journal}{Physical Review B}}
  \textbf{\bibinfo{volume}{86}}, \bibinfo{pages}{220501}
  (\bibinfo{year}{2012}).

\bibitem{shein2013}
\bibinfo{author}{Shein, I.~R.} \& \bibinfo{author}{Ivanovskii, A.~L.}
\newblock \bibinfo{title}{Electronic band structure and fermi surface for new
layered superconductor LaO$_{0.5}$F$_{0.5}$BiS$_2$ in comparison with parent phase
  LaOBiS$_2$ from first principles}.
\newblock \emph{\bibinfo{journal}{JETP letters}} \textbf{\bibinfo{volume}{96}},
  \bibinfo{pages}{769--774} (\bibinfo{year}{2013}).

\bibitem{bjorkman2012}
\bibinfo{author}{Bj{\"o}rkman, T.},
  \bibinfo{author}{Gulans, A.},
  \bibinfo{author}{Krasheninnikov, A.~V.},
  \& \bibinfo{author}{Nieminen, R.~M.}
  \newblock \bibinfo{title}{van der Waals bonding in layered compounds from advanced density-functional first-principles calculations}.
\newblock \emph{\bibinfo{journal}{Physical Review Letters}}
  \textbf{\bibinfo{volume}{108}}, \bibinfo{pages}{235502}
  (\bibinfo{year}{2012}).

\bibitem{wei2013}
\bibinfo{author}{Wei, P.} \emph{et~al.}
\newblock \bibinfo{title}{Exchange-coupling-induced symmetry breaking in
  topological insulators}.
\newblock \emph{\bibinfo{journal}{Physical review letters}}
  \textbf{\bibinfo{volume}{110}}, \bibinfo{pages}{186807}
  (\bibinfo{year}{2013}).

\bibitem{wang2015}
\bibinfo{author}{Wang, Z.}, \bibinfo{author}{Tang, C.}, \bibinfo{author}{Sachs,
  R.}, \bibinfo{author}{Barlas, Y.} \& \bibinfo{author}{Shi, J.}
\newblock \bibinfo{title}{Proximity-induced ferromagnetism in graphene revealed
  by the anomalous hall effect}.
\newblock \emph{\bibinfo{journal}{Physical Review Letters}}
  \textbf{\bibinfo{volume}{114}}, \bibinfo{pages}{016603}
  (\bibinfo{year}{2015}).

\bibitem{checkelsky2014}
\bibinfo{author}{Checkelsky, J.} \emph{et~al.}
\newblock \bibinfo{title}{Trajectory of the anomalous Hall effect towards the quantized state in a ferromagnetic topological insulator}.
\newblock \emph{\bibinfo{journal}{Nature Physics}}
  \textbf{\bibinfo{volume}{10}}, \bibinfo{pages}{731--736}
  (\bibinfo{year}{2014}).

\bibitem{kamihara2008}
\bibinfo{author}{Kamihara, Y.}, \bibinfo{author}{Watanabe, T.},
  \bibinfo{author}{Hirano, M.} \& \bibinfo{author}{Hosono, H.}
  \newblock \bibinfo{title}{Iron-based layered superconductor La[O$_{1-x}$ F$_x$] FeAs   (x= 0.05-0.12) with T$_c$= 26 K}.
\newblock \emph{\bibinfo{journal}{Journal of the American Chemical Society}}
  \textbf{\bibinfo{volume}{130}}, \bibinfo{pages}{3296--3297}
  (\bibinfo{year}{2008}).

\bibitem{kresse1996A}
\bibinfo{author}{Kresse, G.} \& \bibinfo{author}{Furthm{\"u}ller, J.}
\newblock \bibinfo{title}{Efficiency of ab-initio total energy calculations for
  metals and semiconductors using a plane-wave basis set}.
\newblock \emph{\bibinfo{journal}{Computational Materials Science}}
  \textbf{\bibinfo{volume}{6}}, \bibinfo{pages}{15--50} (\bibinfo{year}{1996}).

\bibitem{kresse1996B}
\bibinfo{author}{Kresse, G.} \& \bibinfo{author}{Furthm{\"u}ller, J.}
\newblock \bibinfo{title}{Efficient iterative schemes for ab initio
  total-energy calculations using a plane-wave basis set}.
\newblock \emph{\bibinfo{journal}{Physical Review B}}
  \textbf{\bibinfo{volume}{54}}, \bibinfo{pages}{11169} (\bibinfo{year}{1996}).

\bibitem{blochl1994}
\bibinfo{author}{Bl{\"o}chl, P.~E.}
\newblock \bibinfo{title}{Projector augmented-wave method}.
\newblock \emph{\bibinfo{journal}{Physical Review B}}
  \textbf{\bibinfo{volume}{50}}, \bibinfo{pages}{17953} (\bibinfo{year}{1994}).

\bibitem{perdew1996}
\bibinfo{author}{Perdew, J.~P.}, \bibinfo{author}{Burke, K.} \&
  \bibinfo{author}{Ernzerhof, M.}
\newblock \bibinfo{title}{Generalized gradient approximation made simple}.
\newblock \emph{\bibinfo{journal}{Physical review letters}}
  \textbf{\bibinfo{volume}{77}}, \bibinfo{pages}{3865} (\bibinfo{year}{1996}).

\bibitem{blaha2001}
\bibinfo{author}{Blaha, P.}, \bibinfo{author}{Schwarz, K.},
  \bibinfo{author}{Madsen, G.}, \bibinfo{author}{Kvasnicka, D.} \&
  \bibinfo{author}{Luitz, J.}
\newblock \bibinfo{title}{WIEN2K, An augmented plane wave+ local orbitals
  program for calculating crystal properties},
\newblock \emph{\bibinfo{journal}{Vienna University of Technology,
Austria}}  (\bibinfo{year}{2001}).

\end{thebibliography}

\begin{thebibliography}{10}
\expandafter\ifx\csname url\endcsname\relax
  \def\url#1{\texttt{#1}}\fi
\expandafter\ifx\csname urlprefix\endcsname\relax\def\urlprefix{URL }\fi
\providecommand{\bibinfo}[2]{#2}
\providecommand{\eprint}[2][]{\url{#2}}


\bibitem{tanryverdiev1995}
\bibinfo{author}{Tanryverdiev, V.~S.}, \bibinfo{author}{Aliev, O.~M.} \&
  \bibinfo{author}{Aliev, I.~I.}
\newblock \bibinfo{title}{Synthesis and physicochemical properties of LnBiOS$_2$}.
\newblock \emph{\bibinfo{journal}{Inorganic materials}}
  \textbf{\bibinfo{volume}{31}}, \bibinfo{pages}{1497--1498} (\bibinfo{year}{1995}).

\bibitem{kabbour2006}
\bibinfo{author}{Kabbour, H.} \&
  \bibinfo{author}{Cario, L.}
\newblock \bibinfo{title}{Ae$_2$Sb$_2$X$_4$F$_2$ (Ae= Sr, Ba): New Members of the Homologous Series Ae$_2$M$_{1+n}$X$_{3+ n}$F$_2$ Designed from Rock Salt and Fluorite 2D Building Blocks}.
\newblock \emph{\bibinfo{journal}{Inorganic chemistry}}
  \textbf{\bibinfo{volume}{45}}, \bibinfo{pages}{2713--2717} (\bibinfo{year}{2006}).


\bibitem{guittard1984}
\bibinfo{author}{Guittard, M.} \emph{et~al.}
\newblock \bibinfo{title}{Oxysulfides and oxyselenides in sheets, formed by a rare earth element and a second metal}.
\newblock \emph{\bibinfo{journal}{Journal of Solid State Chemistry}}
  \textbf{\bibinfo{volume}{51}}, \bibinfo{pages}{227--238} (\bibinfo{year}{1984}).


\bibitem{bjorkman2012}
\bibinfo{author}{Bj{\"o}rkman, T.}, \bibinfo{author}{Gulans, A.}, \bibinfo{author}{Krasheninnikov, A.~V.} \&
  \bibinfo{author}{Nieminen, R.~M.}
\newblock \bibinfo{title}{van der Waals bonding in layered compounds from advanced density-functional first-principles calculations}.
\newblock \emph{\bibinfo{journal}{Physical review letters}}
  \textbf{\bibinfo{volume}{108}}, \bibinfo{pages}{235502} (\bibinfo{year}{2012}).

\bibitem{novoselov2005two}
\bibinfo{author}{Novoselov, K.~S.} \emph{et~al.}
\newblock \bibinfo{title}{Two-dimensional atomic crystals}.
\newblock \emph{\bibinfo{journal}{Proceedings of the National Academy of Sciences of the United States of America}}
  \textbf{\bibinfo{volume}{102}}, \bibinfo{pages}{10451--10453} (\bibinfo{year}{2005}).


\bibitem{coleman2011two}
\bibinfo{author}{Coleman, J.~N.} \emph{et~al.}
\newblock \bibinfo{title}{Two-dimensional nanosheets produced by liquid exfoliation of layered materials}.
\newblock \emph{\bibinfo{journal}{Science}}
  \textbf{\bibinfo{volume}{331}}, \bibinfo{pages}{568--571} (\bibinfo{year}{2011}).


\bibitem{yildirim2013}
\bibinfo{author}{Yildirim, T.}
\newblock \bibinfo{title}{Ferroelectric soft phonons, charge density wave instability, and strong electron-phonon coupling in BiS$_2$ layered superconductors: A first-principles study}.
\newblock \emph{\bibinfo{journal}{Physical Review B}}
  \textbf{\bibinfo{volume}{87}}, \bibinfo{pages}{020506} (\bibinfo{year}{2013}).


\bibitem{liu2013}
\bibinfo{author}{Liu, Q.}, \bibinfo{author}{Guo, Y.} \&
  \bibinfo{author}{Freeman, A.~J.}
\newblock \bibinfo{title}{Tunable Rashba effect in two-dimensional LaOBiS$_2$ films: Ultrathin candidates for spin field effect transistors}.
\newblock \emph{\bibinfo{journal}{Nano letters}}
  \textbf{\bibinfo{volume}{13}}, \bibinfo{pages}{5264--5270} (\bibinfo{year}{2013}).


\bibitem{souvatzis2008}
\bibinfo{author}{Souvatzis, P.}, \bibinfo{author}{Eriksson, O.}, \bibinfo{author}{Katsnelson, M.~I.} \&
  \bibinfo{author}{Rudin, S.~P.}
\newblock \bibinfo{title}{Entropy driven stabilization of energetically unstable crystal structures explained from first principles theory}.
\newblock \emph{\bibinfo{journal}{Physical review letters}}
  \textbf{\bibinfo{volume}{100}}, \bibinfo{pages}{095901} (\bibinfo{year}{2008}).
\end{thebibliography}
\end{document}